\documentclass{article}

\usepackage{arxiv}

\usepackage[utf8]{inputenc} 
\usepackage[T1]{fontenc}    
\usepackage{hyperref}       
\usepackage{url}            
\usepackage{booktabs}       
\usepackage{amsfonts}       
\usepackage{nicefrac}       
\usepackage{microtype}      
\usepackage{graphicx}
\usepackage{doi}
\usepackage{blindtext}
\usepackage{enumitem}
\usepackage{epsfig}
\usepackage{graphicx}
\usepackage{amsmath}
\usepackage{amssymb}
\usepackage{multicol}
\usepackage{float}
\usepackage{mcite}
\usepackage[euler]{textgreek}
\usepackage{subcaption}
\usepackage{physics}
\usepackage{siunitx}
\usepackage{cite}

\DeclareSIUnit\molar{\mole\per\cubic\deci\metre}
\DeclareSIUnit\Molar{\textsc{m}}

\setitemize[0]{leftmargin=*}
\setenumerate[0]{leftmargin=*}

\renewcommand{\headeright}{}
\renewcommand{\undertitle}{}
\renewcommand{\shorttitle}{Decoupling environmental modes from tunneling electrons}

\hypersetup{
pdftitle={},
pdfsubject={},
pdfauthor={},
pdfkeywords={},
}

\begin{document}
\title{decoupling environmental modes from tunneling electrons in a partially wet phase molecular mechanically controllable break junction}

\author{C.J. Muller\thanks{Correspondence: PO Box 60, 6580 AB Malden, The Netherlands}
\\ \\
This research has, in its entirety, been privately conducted and funded by the author.
}
\maketitle

\begin{abstract}
A quantum effect at ambient conditions is presented. A benzene dithiol (BDT) molecule or a tetrahydrofuran (THF) molecule is used as a barrier molecule bridging the gold electrodes from a Mechanically Controllable Break junction. It has been known for a long time that the environment of a junction couples to the conduction. With a microscopic layer of fluid it is actually possible to influence this coupling. The electrode-molecule-electrode configuration is lined with a microscopic layer of THF fluid, also known as the “partially wet phase”. This is effectively decoupling the electrode-molecule-electrode system from environmental modes outside of this system. The effects on coherence and quantum interference are exceptional.
\end{abstract}

\begin{multicols}{2}

\section{Introduction}
More than 60 years ago von Hippel \mcite{vonhippel1956source1a,*vonhippel1959source1b} laid out his vision where he saw a world using the intrinsic properties of atoms and molecules: 

"One could build materials from their atoms and molecules for the purpose at hand ... . He can play chess with elementary particles according to their prescribed rules until engineering solutions become apparent."

It was a time where radio tubes were being replaced by integrated circuits and transistors. In contrast to the logical and, already in those days, destined successful path of ongoing miniaturization for IC’s, von Hippel was advocating a bottom up approach which he called Molecular Engineering. With this view von Hippel laid the foundation for a field which we now call Molecular Electronics \mcite{scheer2017molecularelectronics,*datta1997electronictransport,*datta2005quantumtransport}.

During that same time frame, although seemingly unrelated at the time, Gorter \cite{gorter1951source2} established an important finding on the resistivity-temperature curve of granular Sn particle films:

"The granularity of the transported charge manifests itself in the conductance as a result of the Coulomb repulsion of individual electrons. The transfer by tunneling of one electron between two initially neutral regions, of mutual capacitance $C$, increases the electrostatic energy of the system by an amount of $e^2/2C$. At low temperatures and small applied voltages, conduction is suppressed because of the charging energy."

This phenomenon is now known as the Coulomb blockade of single-electron tunneling \cite{likharev1988source3}. Although the physics of granular films and systems need to account for a distribution in particle-size and thus capacitance, theoretical elements used in the early descriptions \mcite{gorter1951source2,zeller1969tunneling,*zeller1968superconductivity,lambe1969chargequantization} are still present in today’s theory for more well defined systems for example a single junction.

A single small-capacitance junction, consisting of superconducting material, was predicted to show a new type of voltage oscillation in 1975 \cite{kulik1975kinetic} transposing the current and voltage character as compared to the normal AC Josephson effect. This is reminiscent of Bloch electrons in a perfect crystal with lattice constant $a$, moving through the energy band representation and being reflected by the Brillouin zone at $2\pi/a$ to the opposing Brillouin zone at $-2\pi/a$, physically identical in nature. A conduction electron under the influence of a constant electric field $E$, will start to oscillate in real space, as a result of interaction with the periodic lattice potential with a Bloch frequency: $f_B=aeE/(2\pi\hbar)$. In real life this effect is not observed in crystals, as the conduction electrons are scattered before they can complete the cycle in the Brillouin zone in $k$ space.

In superconducting small-capacitance current biased junctions the elementary frequency equals $f_{SC}=I/(2e)$, with a corresponding band-zone edge at $Q=+e$ or $-e$ in the energy band representation, $Q$ being the charge on the capacitor. The origin of these predicted voltage oscillations, is the continuous charging until the level reaches $+e$ or $-e$ after which a tunnel event occurs and the recharging starts again. At the tunnel event the system jumps from the band-zone edge to the opposing zone edge, both physically identical in nature, and the cycle starts again. 

The charging of the capacitor with a continuous charge $Q$, smaller than a $2e$ Couper pair charge is physically possible. A moving charge carrier can create some influence charge far smaller as compared to the unit charge of a charge carrier on the small capacitor while moving to the location where it will eventually tunnel, hence treating $Q$ as a classical variable.

The interest in this system originates from the uncertainty relation between $\varphi$, the superconductor order parameter, and $Q[\varphi,Q]=i2e$. When $\varphi$ is well defined, the system is described by the AC Josephson effect, it implies that $Q$ is undefined, which is equivalent to a large capacitance system. In the other extreme, for small capacitance systems, theoretically $Q$ is well defined and $\varphi$ is undefined, leading to the small-capacitance effects as described above. 

In the mid 1980’s several authors of the first hour \cite{benjacob1985ONE,ben1988TWO,averin1985ONE,guinea1986coherent} theoretically analyzed small-capacitance effects in both normal metal and superconducting metal junctions, they predicted that both the superconducting as well as the normal metal small capacitance junction would each have their own type of voltage oscillation. The normal metal type single electron tunneling (SET) oscillations will have a frequency $f_N=I/e$, the tunnel event occurs at the zone edge $+e/2$ or $-e/2$ in the energy band representation. A rigorous theoretical underpinning of these predicted effects \cite{averin1986TWO,likharev1985theory} initiated a widespread experimental research, for extensive early reviews see \cite{averin1991THREE,schon1990ONE}. Defining a phase $\varphi_S(t)=(e/\hbar)\int^{t}_{-\infty}V(t')\dd t'$ being the integral over the voltage across the junction, Schön \cite{schon1985TWO} demonstrated the commutation rule $[\varphi_S, Q]=ie$ for normal metal junctions. It was well recognized that with a small-capacitance junction we might have a system with the possibility to complete the full cycle through the energy band representation and arrive at a novel quantum effect.
  
Experiments on small capacitance single junctions became possible due to advanced e-beam lithography capabilities in the 80’s. A capacitance of the order of \SI{e-15}{F} leads to a charging energy equivalent to \SI{1}{\kelvin}, a temperature well accessible by cryogenic techniques. Results showed that the direct electrical environment of the junction played a crucial role in the occurrence of a Coulomb blockade gap in the current-voltage (IV)-curve. Parasitic capacitance of the electrical leads connecting to the junction is in the \si{pF} range, much larger than the junction capacitance, leading to a high capacitance-value as seen by the junction and thus easily destroying any $E_C=e^2/2C$ charging-energy effects. This was prompting experiments where the junction was decoupled from the leads by inclusion of large resistors, with values much larger than $R_Q=h/e^2$ in the electrical leads at very close proximity, \si{\micro\meter}’s, to the junction \cite{cleland1990charge}. 

An alternative elegant way to decouple a small capacitance from the leads, is by implementing two high resistive tunnel junctions, with tunnel resistances $R_T\gg R_Q$, on either side of a conductive island \cite{fulton1987observation}, very similar to Gorter’s use of granular films where small Sn particles are connected via oxidized tunnel barriers. These structures are known as "dots". The low capacitance central island part of the dot is perfectly isolated from environmental modes in the leads by the two high impedance tunnel junctions. The coupling or tunnel rates of these model systems to the leads can be tuned by the characteristic resistance of the tunnel junctions. The capacitance can be low as it scales with the size of the dot. The condition $R_T\gg R_Q$ for the two tunnel junctions ensures that after tunneling to the dot, the electron is localized there. The dot system thus exhibits both the wave character, tunneling to and from the dot, as well as the particle character of an electron as there can only be an integer number of electrons on the dot. If $R_T$ becomes smaller than $R_Q$ the charge becomes delocalized along the system and charging effects will be suppressed.

There are a number of dot regimes possible \cite{kouwenhoven1997electron}. A metallic grain of \si{\micro\meter} size holds billions of electrons occupying billions of states up to the Fermi energy. The coulomb repulsion by putting 1 additional electron on this grain requires this electron to have an excess energy of $e^2/2C$, before being able to enter the grain. The Fermi wavelength is of the order of the inter atomic distance, much smaller as the particle size, ensuring the metal electron states are as in bulk metal, finely dispersed with energy separations much smaller than the $e^2/2C$ value.

A different regime holds in a semiconductor dot also of \si{\micro\meter} dimensions. The Fermi wavelength in semiconductors is of the order of \SI{100}{\nano\meter}. Now the charging energy as well as finite size effects in this dot interact with one-another. When the charging energy of these dots become comparable to the energy level spacing it is predicted that the spin degeneracy is lifted and a regular renormalized level spacing is the resultant \cite{beenakker1991granular}. Because next to the charging effects the finite size effects of these types of dots start to become a crucial factor in the description these dots are called quantum dots.

Where Gorter meets von Hippel we arrive at the ultimate quantum dot. One molecule attached via two tunnel junctions to the leads. Note that this dot diameter is smaller by more than a factor of 1000, as compared to the quantum dot regime detailed above.  HOMO and LUMO orbitals are interacting with the charging energy and are expected to play a role in the conduction. Free molecules have a discrete set of electronic levels, typical level spacing is in the \si{\electronvolt} range. Molecules exhibit vibrational and rotational degrees of freedom, where the typical vibrational energies are in the \SI{0.1}{\electronvolt} range and the rotational energies are in the \si{\milli\electronvolt} range. One of the early and most spectacular successes of quantum theory is the answers it provided to questions related to molecular structure and spectra. Now we are faced with questions related to a single connected molecule. Will such a system still be described by an electrostatic capacitance? As the molecule only measures a few \si{\angstrom} it is not safe to assume the screening length in the molecule to be smaller as compared to the size of the molecule. It cannot be assumed to be an electrostatic capacitance. For simplicity, this system is treated as if it can be described by a single electrostatic capacitance value.

In all of the above described systems, the environment of the junction has played an important role ever since the first experiments on small capacitance junctions. It is important to define the quantum mechanical environment \cite{devoret1992source4a,ingold1992source4b,devoret1990effect}. Next to the system defining modes, determined for example by junction capacitance, inductance, resistance, the tunneling electron can couple to other modes. These modes can couple to the junction via the leads (lead environmental modes) or via the vacuum (vacuum environmental modes). The vacuum environmental modes reflect the coupling between tunneling electrons and modes in the medium enclosing the electrode-molecule-electrode configuration. These modes are present in every enclosing medium, even in vacuum, hence “vacuum environmental modes” even though the enclosing medium may technically not be a vacuum. 

In reality the different types of modes are hard to distinguish, a certain environment encompasses a coupling of various modes and mode types \cite{devoret1992source4a,ingold1992source4b,devoret1990effect}. Quantum dots with $R_T\gg R_Q$ are well decoupled from lead environmental modes. The decoupling from vacuum environmental modes is however from an experimental point of view difficult to influence. It is important to realize that a tunneling electron can couple to vacuum environmental modes and loss or gain energy to or from these modes. With increasing frequency, the coupling to vacuum environmental modes becomes easier as high frequency processes experience very low impedance in vacuum, as the impedance of vacuum equals \SI{377}{\ohm} only, far less than $R_Q$.

A study of partially wet phase molecular junctions is presented, where a molecule is bridging the gap between two electrodes of a MCB junction. In the ideal situation, the system defining modes are the only relevant modes for tunneling electrons to couple to. Ideally all lead environmental modes and vacuum environmental modes are decoupled from the electrode-molecule-electrode system. Below it will be shown that nature provides for such a system.

\section{Experimental Setup and Way of Work}
\begin{figure}[H]
    \centering
    \includegraphics[width=.45\textwidth]{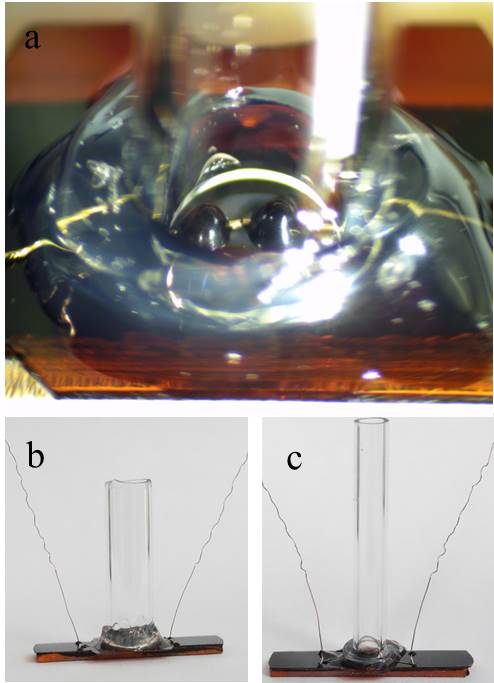}
    \caption{Bending beam assemblies with mounted cells, the beam measures $1\times5\times24$ \si{\milli\meter}. Photo a and c show the $\phi_{\text{in}}^{2}$ cell with a typical length of \SI{26}{\milli\meter} . The inside of the cell shows the junction in the center, the liquid rubber is applied from the outside ensuring a tight and flexible seal. Photo b shows the $\phi_{\text{in}}^{4.5}$ cell with a typical length of \SI{22}{\milli\meter}. Copper leads complete the connections.}
    \label{fig:1}
\end{figure}

Experiments are executed at ambient conditions. The classical notched filament MCB junction \mcite{muller1992source1a,*muller1992source1b,*muller1992source1c} has been used. A phosphor bronze bending beam is laminated with a \SI{200}{\micro\meter} thick insulating Kapton layer, glued with Stycast epoxy to the bending beam. Filament material is either made up of 99.99 \si{\percent} hard temper Au ($\text{Au}^{\text{H}}$) or \SI{99.99}{\percent} soft temper Au ($\text{Au}^{\text{S}}$) glued with Stycast epoxy to the Kapton insulated bending beam. A glass liquid cell is mounted surrounding the junction area using a liquid rubber. Fig.~\ref{fig:1} shows the bending beam assembly. Two types of cells have been used. One with a \SI{4.5}{\milli\meter} inner diameter ($\phi_{\text{in}}^{4.5}$), which releases the solution upon pivoting of the setup to the cell opening, where the solution can be absorbed. The other cell inner diameter measures \SI{2}{\milli\meter} ($\phi_{\text{in}}^{2}$). This cell needs to be drained with a paper tip due to the capillary forces holding the fluid in the cell upon tilting.

In the measurements no dependency could be detected on the temper of the gold filament. Some dependency was detected related to the inner diameter of the cell, $\phi_{\text{in}}^{2}$ or $\phi_{\text{in}}^{4.5}$. For completeness the temper of the gold and the inner cell diameter are listed for reported data of every bending beam assembly.

The experimental setup makes use of a pivoting MCB setup, enabling the partially wet phase \cite{bonn2009wetting}, as detailed in ref \cite{muller2021pivoting}. In this setup the MCB bending beam assembly can be pivoted to a desired angle, to either submerge the junction in fluid, or to drain the junction area from fluid while the adjusted junction remains intact during pivoting. For bending beam assemblies using the $\phi_{\text{in}}^{4.5}$ cell, the pivoting was used to drain the cell, after draining the cell was placed in vertical position again. For bending beam assemblies using the $\phi_{\text{in}}^{2}$ cell the setup was maintained in a fixed vertical cell position.

The IV curves are recorded with a HP4155 Semiconductor Parameter Analyzer. The junctions are voltage biased, one side of the junction is kept at zero the other side was voltage scanned, while measuring the current. The HP4155 exhibits 4 independent signal monitoring units (SMU’s) of which 2 are required for a measurement. The reproducibility of the obtained results has been validated on 2 independent sets of SMU’s. The HP4155/measurement setup attached to the MCB setup has been regularly tested for linearity with a \SI{50}{\mega\ohm} resistor connected at the device location.

Double shielded cables are used to connect the HP4155 to the device; the last few centimeters to the junction are unshielded. No additional filtering is used. The piezo HV supply, normally in use for fine tuning of a MCB junction, is not used for the reported experiments. IV traces (single direction, indicated by small arrows) consist of 1000 data points, recorded in 45 seconds. The time between consecutive data points remains \SI{45}{\milli\second} where partial traces are shown. The scans which are recorded faster are indicated. On some occasions two consecutive scans with opposing scan direction are shown, in these cases also each scan consists of 1000 data points and there is no delay at the point of scan reversal.

In the 90‘s Tour was pioneering the synthesis of di-thiolate molecules \cite{tour1995self} which exhibit 2 opposing thiol anchor groups, enabling the initial MCB BDT experiments at Yale \cite{mullervleeming1996atomicprobes}. In the below experiments commercially available BDT molecules (99+ \si{\percent} pure) are dissolved in 99.85+ \si{\percent} anhydrous ($<50$ \si{ppm} $\text{H}_2\text{O}$) tetrahydrofuran stabilized with 200-250 \si{ppm} butylated hydroxyl-toluene. 

Two different types of experiments have been performed. In one type a BDT THF solution is used with a concentration of approximately \SI{1}{\milli\Molar}, where the intent is to create a THF partially wet phase electrode-BDT molecule-electrode structure. The other experiment type made use of the pure THF fluid with the intention to create a THF partially wet phase electrode-THF molecule-electrode structure. In each type of experiment the solution/fluid is injected into the cell prior to breaking the filament. The molecules and THF fluid are handled under inert Argon gas. Standard cannulation techniques for handling air sensitive reagents have been used. Only at injecting the BDT/THF solution or the pure THF fluid into the cell, it is exposed to air.
The following experimental way of work has been followed in the described experiments:
\begin{itemize}
    \item	The MCB setup is pivoted such that the liquid cell is in upright, vertical position. 
    \item	The liquid cell is injected with either the \SI{1}{\milli\Molar} BDT solution or the THF fluid by a glass syringe with stainless needle, while the bending beam assembly filament is still unbroken.
    \item	The motor is switched on while monitoring the conductivity of the filament, once the filament is broken the motor is immediately switched off. Based on the spindle velocity as well as the bending beam assembly attenuation factor it is estimated that the electrode separation at this stage is in the order of nanometers. 
    \item	In this situation the junction is completely immersed in fluid and thus in the completely wet phase. Approximately 2-5 minutes settling time is provided for wetting the newly formed electrodes and for molecules to adhere, prior to draining.
    \item	The bending beam assembly is drained, either by pivoting the setup ($\phi_{\text{in}}^{4.5}$ cell) or by a paper tip ($\phi_{\text{in}}^{2}$ cell). Draining is performed in such a way that it takes about half an hour to get to the completely dry phase.
    \item	Directly after draining the setup is pivoted back to the original position in case the $\phi_{\text{in}}^{4.5}$ cell
    has been used; IV measurements can be recorded while the junction area slowly dries. Alternatively a current measurement can be taken (for example at \SI{-5}{\volt}) every few minutes where full IV measurements are only started once the partially wet phase has been reached.
    \item	Once the setup is in the original position, it is not to be touched any more. The drying process progresses the junction through the following phases: 
    \begin{itemize}
        \item	The completely wet phase, once broken and still immersed in fluid and also just after draining
        \item	The partially wet phase, where a layer thickness of only 1 or a few THF molecules will line the electrode-molecule-electrode system. 
        \item	The completely dry phase, defined as completely dry from THF. 
    \end{itemize}
    \item	After reaching the completely dry phase the MCB junction has to be discarded of and needs to be replaced with a new (unbroken filament) bending beam assembly.
\end{itemize}
A limited number of experiments are carried out in the H$_2$O partially wet phase. In order to arrive at this phase a regular household humidifier is put together with the MCB setup and a hair-hygrometer in an acrylate plastic enclosure.

The majority of the experiments are performed in the THF partially wet phase, limiting measurement time to a few minutes only before entering the completely dry phase. During this time the IV traces still change as a result of the continuous evaporation of THF molecules. Reproducing the results within the same bending beam assembly is therefore inherently not possible. For this reason only results are shown which have been demonstrated in multiple bending beam assemblies. The periodicity of the presented data, the validated data on two SMU sets, the specific reproducing line shape of the oscillations and in some cases the demonstrated progression of the effects over a short time lapse, all should provide confidence in the presented data despite the lack of reproducibility within the same bending beam assembly.

\clearpage

\section{Experimental Results}

\subsection{Field emission oscillations}
\begin{figure}[H]
    \centering
    \includegraphics[width=.49\textwidth]{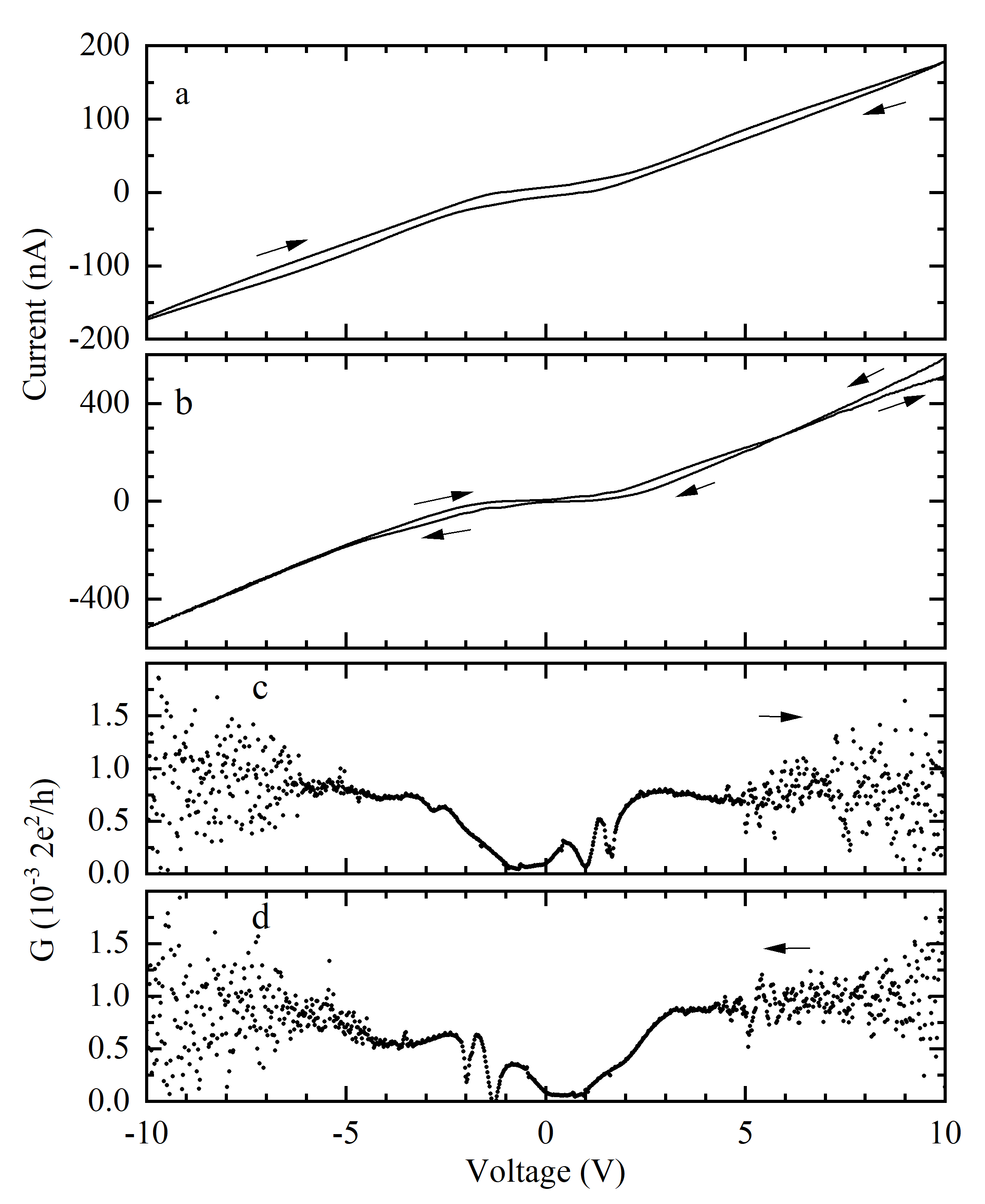}
    \caption{Panel 2a shows a typical IV curve after a junction is broken and immersed in THF fluid, Au\textsuperscript{H}, $\phi_{\text{in}}^{4.5}$. Panel 2b shows data from a bending beam assembly in the drying process after being immersed in a BDT/THF solution, revealing enhanced conductivity and also similarity in shape to the curve shown in panel a, Au\textsuperscript{H}, $\phi_{\text{in}}^{4.5}$. In panel c and d the conductivity of the two traces in panel b are shown. Clear voltammetry behavior is visible.}
    \label{fig:2}
\end{figure}

Once the bending beam assembly filament is broken in pure THF, IV curves are as indicated in Fig.~\ref{fig:2}a. The IV curves show a cyclic voltammetry component, exhibiting a non-zero current at $V=0$. Also structure in the IV curve is observed to reproduce at a voltage value which switches polarity once the scan direction is reversed. Towards higher voltages a more or less linear part shows in the IV curve which exhibits none or only minor hysteresis. 

\begin{figure}[H]
    \centering
    \includegraphics[width=.49\textwidth]{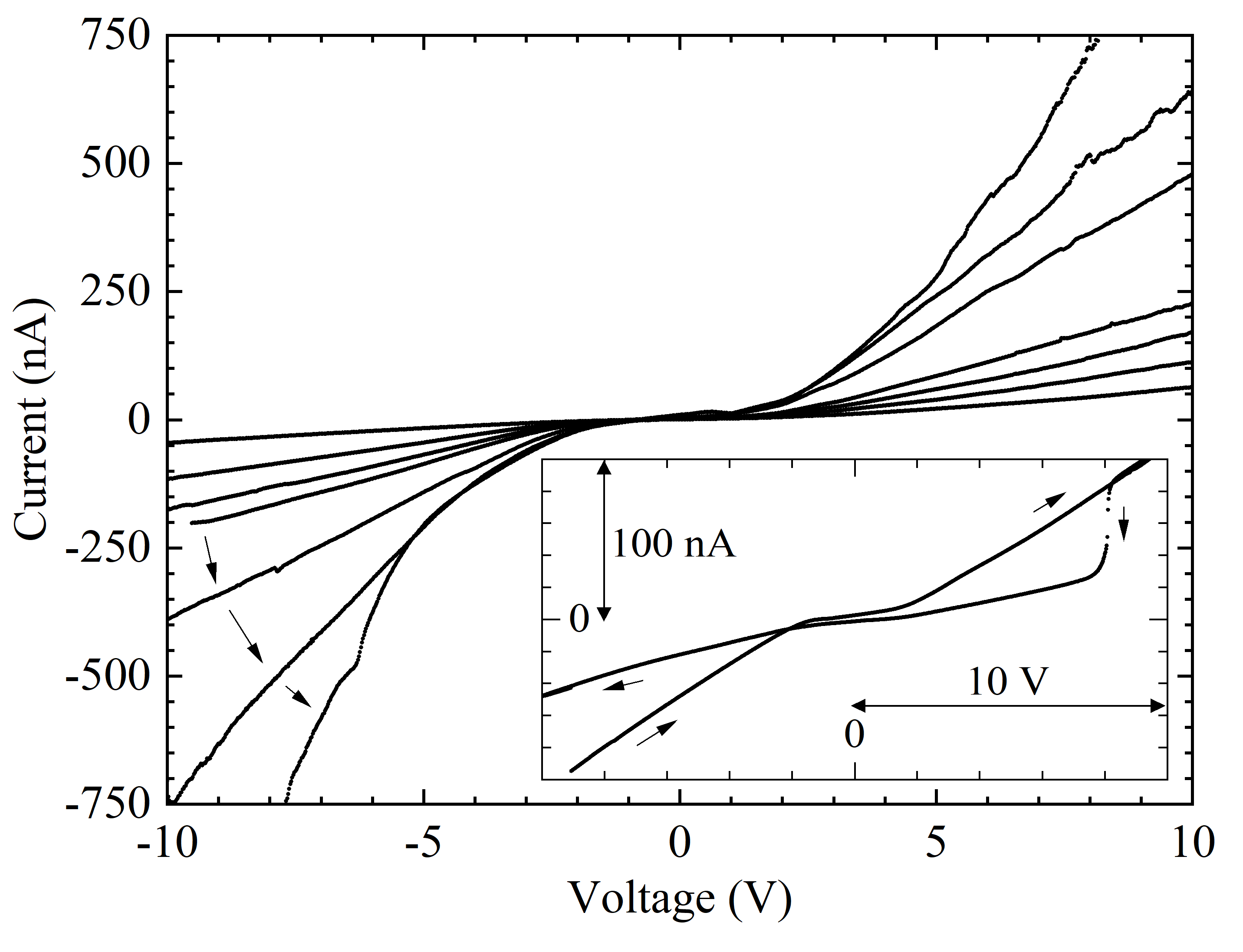}
    \caption{Typical drying behavior of a junction submersed in a BDT/THF solution after the solution has been drained. The arrows indicate the continuous increase in conductivity over time. The inset shows the effect of the draining on the junction; a steep decrease in conduction at the point in time where the cell is drained, Au\textsuperscript{H}, $\phi_{\text{in}}^{4.5}$.}
    \label{fig:3}
\end{figure}

This IV curve is very similar in shape to an IV curve of a junction immersed in a \SI{1}{\milli\Molar} BDT/THF solution of which the conduction at \SI{10}{\volt} is approximately \SI{100}{\nano\ampere} see the inset of Fig.~\ref{fig:3}. These IV curves of devices submerged in fluid are independent on electrode gap distance unlike the exponential influence on the tunnel current for a vacuum gap. In this case we cannot draw any positional conclusion from these curves as to how large the gap between the electrodes actually is. Crashing the electrodes into each other leads to a metallic constriction. The resistance of the smallest possible one atom point-contact amounts to approximately $h/2e^2\approx\SI{12}{\kilo\ohm}$, which would show as a vertical line through the origin on the scale in Fig.~\ref{fig:2}a.

The limits of cyclic voltammetry will be investigated under the continuous reduction of the amount of fluid, until we reach a single layer of fluid molecules, solidly adhered to the electrodes. In Fig.~\ref{fig:3} this is demonstrated during the drying process for a BDT/THF solution. The effect of the draining process can be registered during a measurement, indicated in the inset of Fig.~\ref{fig:3} as a steep decrease of the current. In general a current reduction at draining of about 50-65 \si{\percent} is observed. After draining, the wet electrodes and junction area will slowly start to dry which lasts approximately half an hour. The conductance of the junction increases continuously during the drying process. The IV curves can be recorded during this process, as the duration to take a trace is short as compared to the duration of the drying process. The IV curves in Fig.~\ref{fig:3} are recorded at fixed time intervals for a positive scan direction during the drying process. Just after draining the current is approximately \SI{50}{\nano\ampere} at \SI{10}{\volt}. Increases during the drying process to values of \SI{3}{\micro\ampere} at \SI{10}{\volt} have been measured, prior to a sudden conductivity decrease upon entering the completely dry phase.

A law of nature seems to hold for two electrodes spaced at nanometers distance during the drying process. Be it soft temper Au or hard temper Au used as the MCB electrode material, be it broken in air and after that immersed in fluid or broken directly in a THF environment, the shown electrode behavior upon the drying process is the same. Once the excess fluid has been drained, the continuous increase of the conductivity is attributed to a constantly decreasing distance between the two closest points on the opposing electrodes due to the lasting evaporation of the fluid on and between the electrodes. Apparently there is enough elasticity in the nanometers separated electrodes such that the exact initial separation does not matter. The continuously decreasing amount of fluid between the electrodes pulls the electrodes ever closer together. This process continues until the fluid is getting depleted, and a molecule is squeezed between the electrodes, blocking any further movement. The adhesion forces of the molecules (either BDT or THF) to the gold electrode surface, together with the partially wet phase layer which is lining the entire electrode-molecule-electrode system are such that a metal-molecule-metal junction is favored over a metallic junction.

The above also holds for electrodes broken in air where a metallic microscopic contact is re-established by reversing the motor action and subsequently exposing the junction to the solution. Once the solution is submitted the microscopic metallic contact is “broken” because fluid gets in between the electrodes.

Fig.~\ref{fig:2}b shows data from a bending beam assembly in the drying process after being immersed in a BDT/THF solution. Although the conductivity has increased by a factor of 10 compared to the value just after draining, there is still a large similarity with the curve in \ref{fig:2}a immersed in pure THF. The conductance of the IV traces in \ref{fig:2}b is shown in \ref{fig:2}c and d. Typical voltammetry behavior is observed at the structure between 1-2 \si{\volt} which reproduces in the opposing scan direction at opposite polarity. In addition an increase in noise towards higher voltages shows.

\begin{figure}[H]
    \centering
    \includegraphics[width=.49\textwidth]{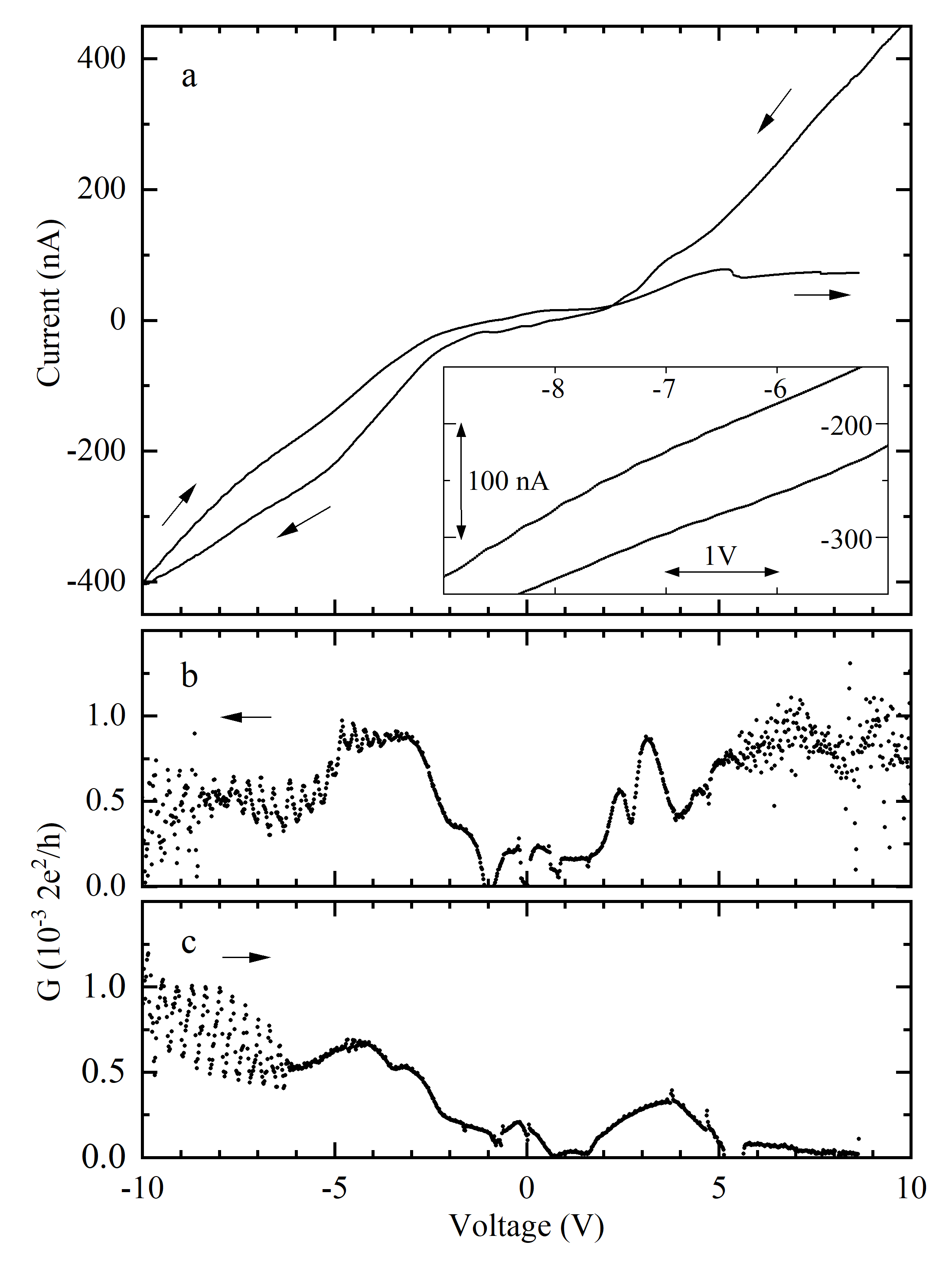}
    \caption{Panel a shows an IV curve in the partially wet phase of a junction submitted to the BDT/THF solution. During the increasing voltage scan the transition to the completely dry phase occurs at positive voltages. Panel b and c present the associated conductance showing clear oscillations at negative voltages. Panel b also shows an increase in noise towards larger voltages, Au\textsuperscript{H}, $\phi_{\text{in}}^{4.5}$.}
    \label{fig:4}
\end{figure}

This is consistent with Fig.~\ref{fig:4}, which shows partially wet phase data of a junction which also has been immersed in a BDT/THF solution. In the conductance data in panel b and c also an increase in noise exists for positive polarity in panel b. Conductance oscillations show in both panels b and c, for negative voltages in excess of \SI{-4}{\volt}. In panel c the conductance oscillations are bounded by a smooth envelope, sometimes called “wave package” or “frequency beating”. Oscillations in the conductance at voltages in excess of the work-function $\Phi^{\text{Au}}$, approximately \SI{5}{\volt} for gold, are well known and are usually explained in the context of field emission resonance \mcite{gundlach1966berechnung,binnig1982helv,kolesnychenko2000field,*kolesnychenko2002experimental}.

The IV curve in the positive scan direction shows a steep decrease in conduction which is typical for entering the completely dry phase. In the completely dry phase the THF is fully evaporated. The remaining molecules on and between the electrodes are strongly adhering to the electrodes and are now exposed to water and air molecules. The current is quickly reduced from a maximum value just prior to entering the completely dry phase to \si{\pico\ampere} values in a matter of minutes. In this situation the electrodes are still in contact with undefined molecules between them acting as glue. After entering the completely dry phase the bending beam assembly can no longer be used as the electrodes are contaminated with undefined molecules and have been exposed to air and water molecules.

\begin{figure}[H]
    \centering
    \includegraphics[width=.49\textwidth]{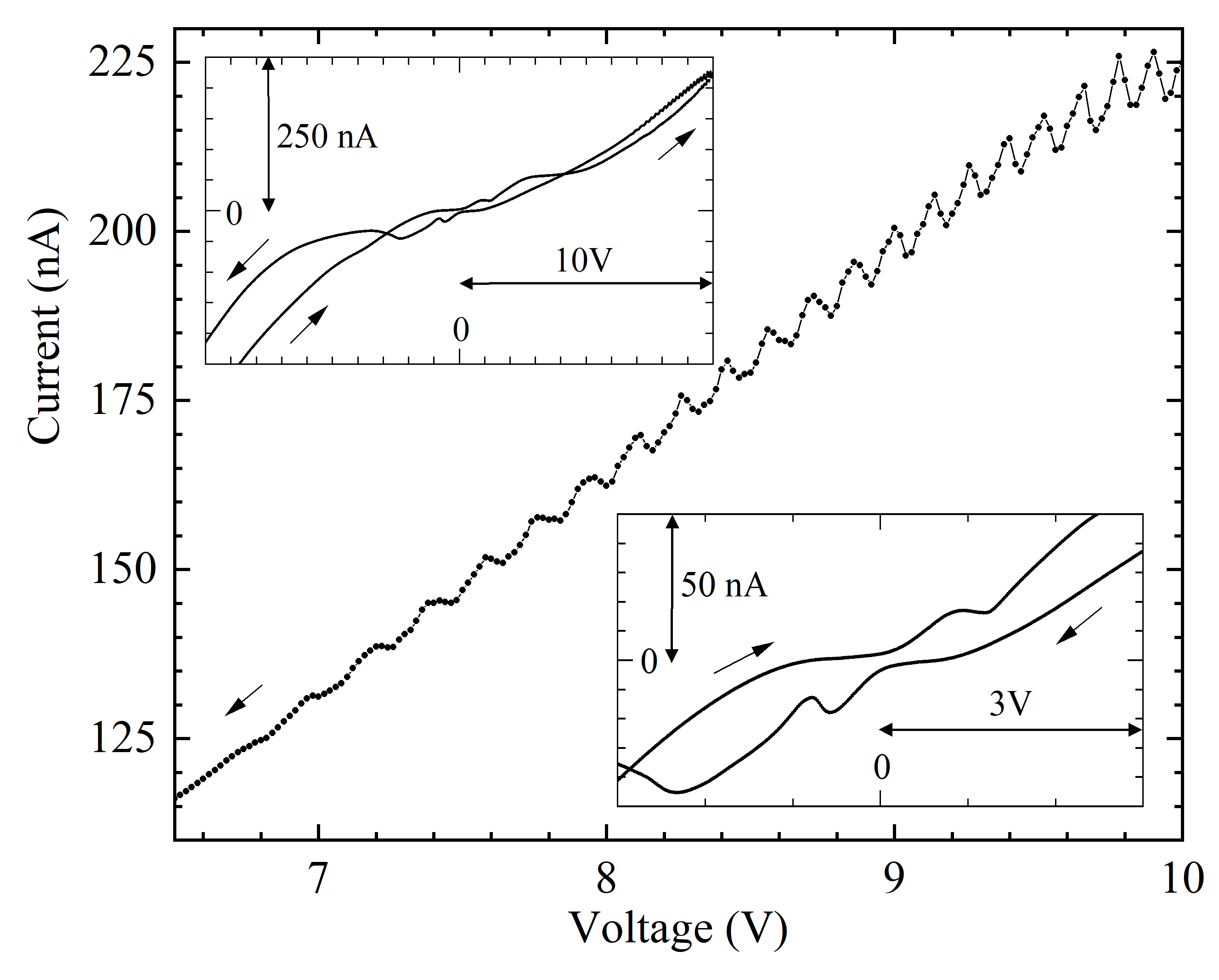}
    \caption{IV curve from a junction in the partially wet phase which has been exposed to the THF fluid. Oscillations in the current show over a substantial voltage range. The upper inset shows the entire IV curve over a \SIrange{-10}{10}{\volt} range. The lower inset shows the central part of the IV curve, Au\textsuperscript{S}, $\phi_{\text{in}}^{2}$.}
    \label{fig:5}
\end{figure}

\begin{figure}[H]
    \centering
    \includegraphics[width=.49\textwidth]{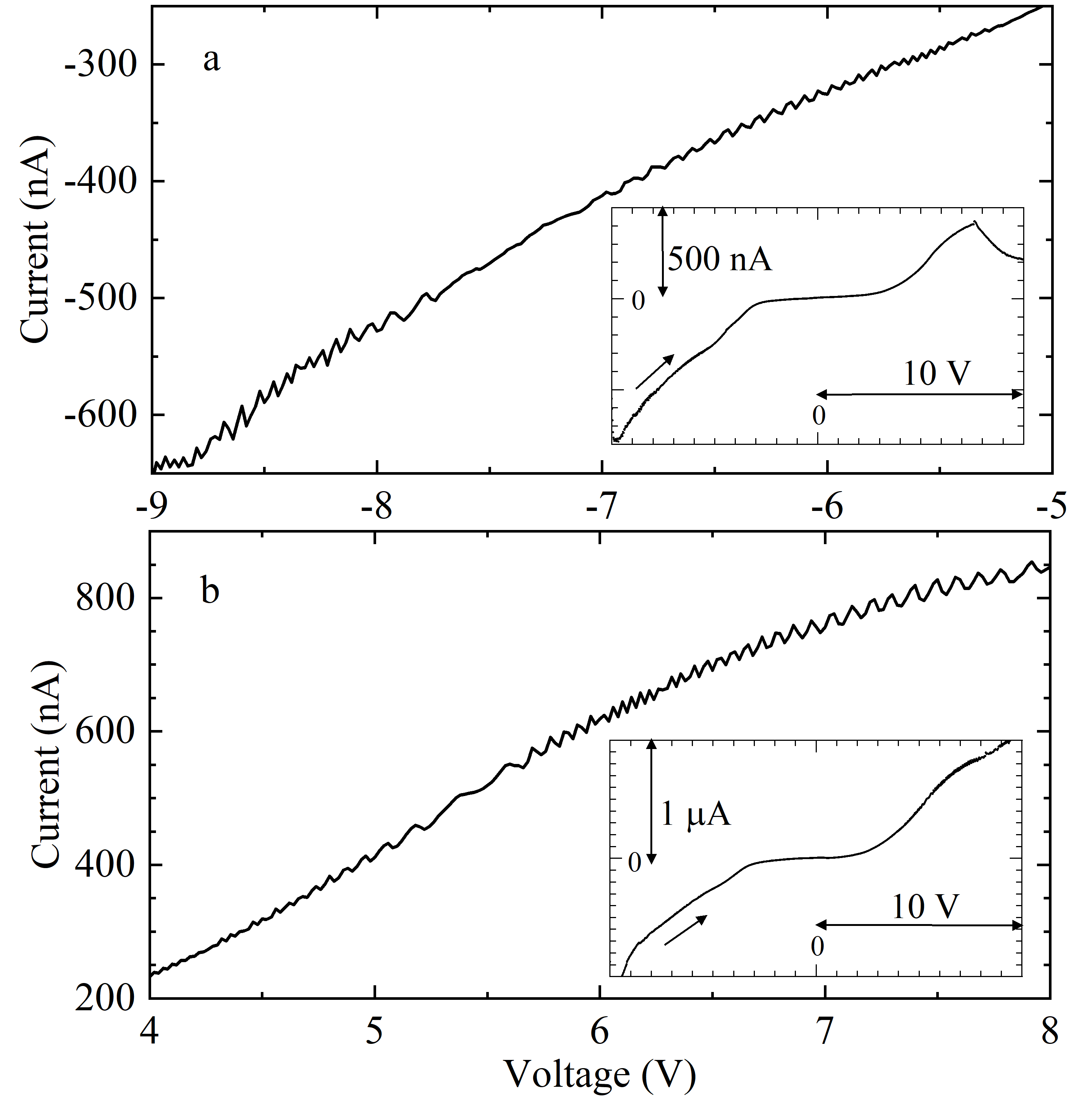}
    \caption{Current oscillations are common in the partially wet phase as shown in panel a and b, where both junctions have been submitted to the BDT/THF solution. The insets in panel a and b show the entire IV curve over the \SIrange{-10}{10}{\volt} range. An abrupt transition to the completely dry phase close to \SI{8}{\volt} is demonstrated by the inset in panel a. Panel a: Au\textsuperscript{S}, $\phi_{\text{in}}^{2}$, panel b: Au\textsuperscript{S}, $\phi_{\text{in}}^{4.5}$.}
    \label{fig:6}
\end{figure}

Fig.~\ref{fig:5} presents data in the partially wet phase from a bending beam assembly which has been exposed to the THF fluid. The IV curve over the full \SIrange{-10}{10}{\volt} range is shown in the upper inset, an enlargement of the central section is shown in the lower inset. Also this IV curve shows pronounced oscillations in this case however the current oscillates, showing in the IV curve as negative differential resistance parts. This is uncommon in devices where the electrodes are made from metal and where the density of states in the electrodes is expected to be fairly constant at the Fermi energy. It is posed that partially wet phase molecular junctions created in the way described above should not be defined by a fairly constant density of states in the electrodes.
Current oscillations can also be observed in Fig.~\ref{fig:6}a and Fig.~\ref{fig:6}b which details IV traces of two different bending beam assemblies in the partially wet phase after being exposed to a THF/BDT solution. The insets show the corresponding large scale IV curve for both bending beam assemblies. Fig.~\ref{fig:6}a and Fig.~\ref{fig:6}b show again “wave package” or “frequency beating” where alternating parts of the IV curve do and do not show current oscillations. 

For Fig.~6a the oscillations occur at negative bias, for Fig.~\ref{fig:6}b they occur at positive bias. In both cases the oscillations are occurring at voltages in excess of $\abs{4}$ \si{\volt}. The inset of Fig.~\ref{fig:6}a shows an abrupt transition to the completely dry phase near \SI{8}{\volt}. Although gradual transitions to the completely dry phase also occur, it shows that defining the point in time where the partially wet phase ends is not difficult. 

Normally the partially wet phase exists just prior to entering the completely dry phase. The transition between an increasing and a decreasing conductance is an indication that the completely dry phase nears. In addition, the partially wet phase in general shows oscillations in the current at larger ($eV>\Phi^{\text{Au}}$) voltages.

\subsection{Oscillations in a THF barrier THF partially-wet-phase junction}
In this section all devices have been immersed in pure THF fluid, after draining and drying the results are recorded in the partially wet phase. The bridging THF molecule(s) between the electrodes are considered constituting the tunnel junction barrier. The THF layer lining the device provides for the partially wet phase.

\begin{figure}[H]
    \centering
    \includegraphics[width=.49\textwidth]{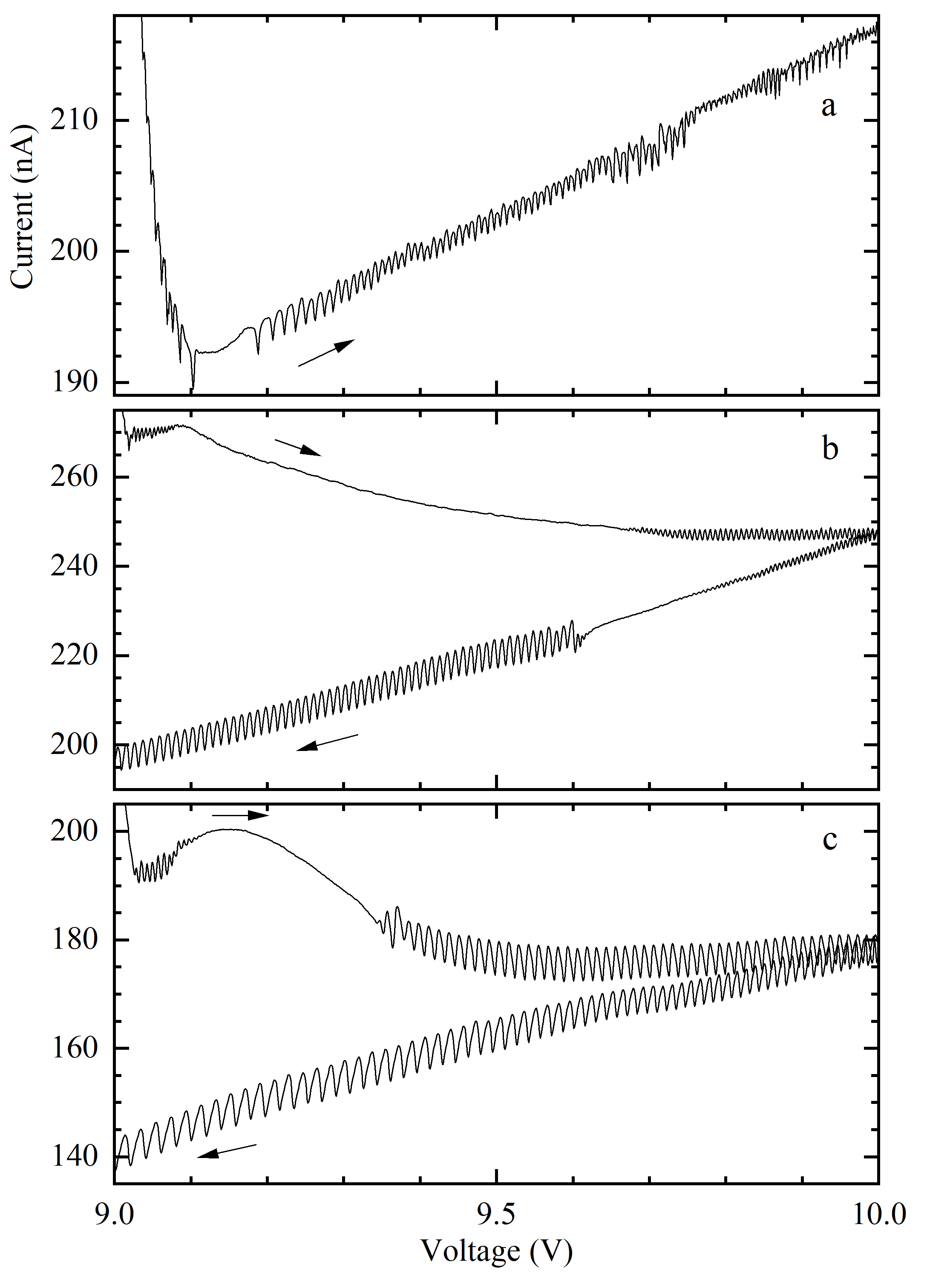}
    \caption{A THF barrier junction in the partially wet phase progressing over time. In panel b oscillations are seen to be developing near \SI{10}{\volt}. In panel c, about 2.5 minutes later, these oscillations are developed at \SI{10}{\volt}, Au\textsuperscript{S}, $\phi_{\text{in}}^{2}$.}
    \label{fig:7}
\end{figure}

The experiments presented in the following sections zoom in at a high voltage offset with a \SI{1}{\milli\volt} resolution for a \SI{1}{\volt} scan and a \SI{2.5}{\milli\volt} resolution for a \SI{2.5}{\volt} scan. Due to the nature of the experiment and the short duration of the partially wet phase no data for the large scale \SIrange{-10}{10}{\volt} total IV curve have been measured. In general the IV curve at large \SIrange{-10}{10}{\volt} scale for THF barrier junctions can be expected to be similar in shape to the one shown in Fig.~\ref{fig:5}. Data from Fig.~\ref{fig:7} panel a, b and c are measured for the same junction with about \SI{1}{\minute} intervals. In panels b and c an increasing and decreasing voltage scan are part of the same scan, in panel a only the increasing scan is shown. The curve in panel a starts with a steep decrease in current, this “switch on” effect is followed by a linear increase. Initially the drying process increased the conductivity going from panel a to panel b as can be observed by the current levels. Going from panel b to panel c the overall conduction has started to decrease. A clear pattern of oscillations is revealed. In a timescale of several minutes the pattern changes with the overall trace changing as a result of the drying process. The amplitude of the oscillations is more or less constant for the larger stretches in panel b and c. Panel c in addition shows that the period of the oscillations is not constant. The lower curve shows periods from \SI{21}{\milli\volt} at \SI{9}{\volt} to \SI{12}{\milli\volt} at \SI{10}{\volt}. In the upper curve the period varies from \SI{14}{\milli\volt} at \SI{9.35}{\volt} to \SI{12}{\milli\volt} at \SI{10}{\volt}. Parts of Fig.~\ref{fig:7} have been enlarged in Fig.~\ref{fig:8}. The line-shapes of the oscillations in the different panels have a sharp “V shaped” minimum in common. The top part of the line-shape is rounded. The line-shape in Fig.~\ref{fig:8}b2 is more or less vertically oriented, approximately 2 minutes later the increasing voltage scan in Fig.~\ref{fig:8}c show line-shapes leaning to the left while the decreasing scan shows line-shapes leaning to the right.

\begin{figure}[H]
    \centering
    \includegraphics[width=.49\textwidth]{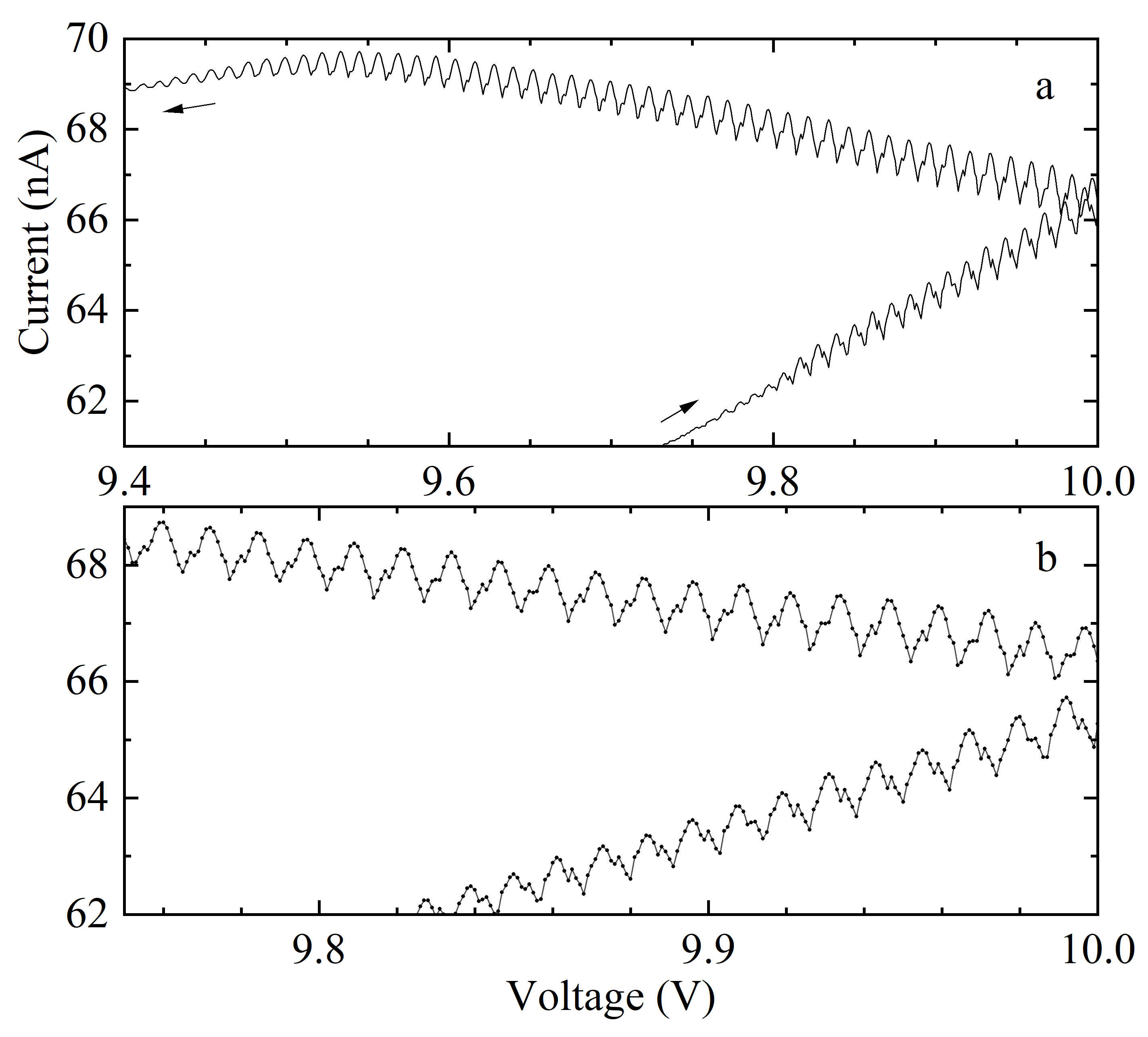}
    \caption{Panel a and b show the “V-shaped” oscillations in the partially wet phase. Structure of only \SI{2}{\milli\volt} wide is seen to reproduce in the oscillations. The lower curve in panel b is offset \SI{-1}{\nano\ampere} for clarity, Au\textsuperscript{S}, $\phi_{\text{in}}^{2}$.}
    \label{fig:9}
\end{figure}

\begin{figure}[H]
    \centering
    \includegraphics[width=.49\textwidth]{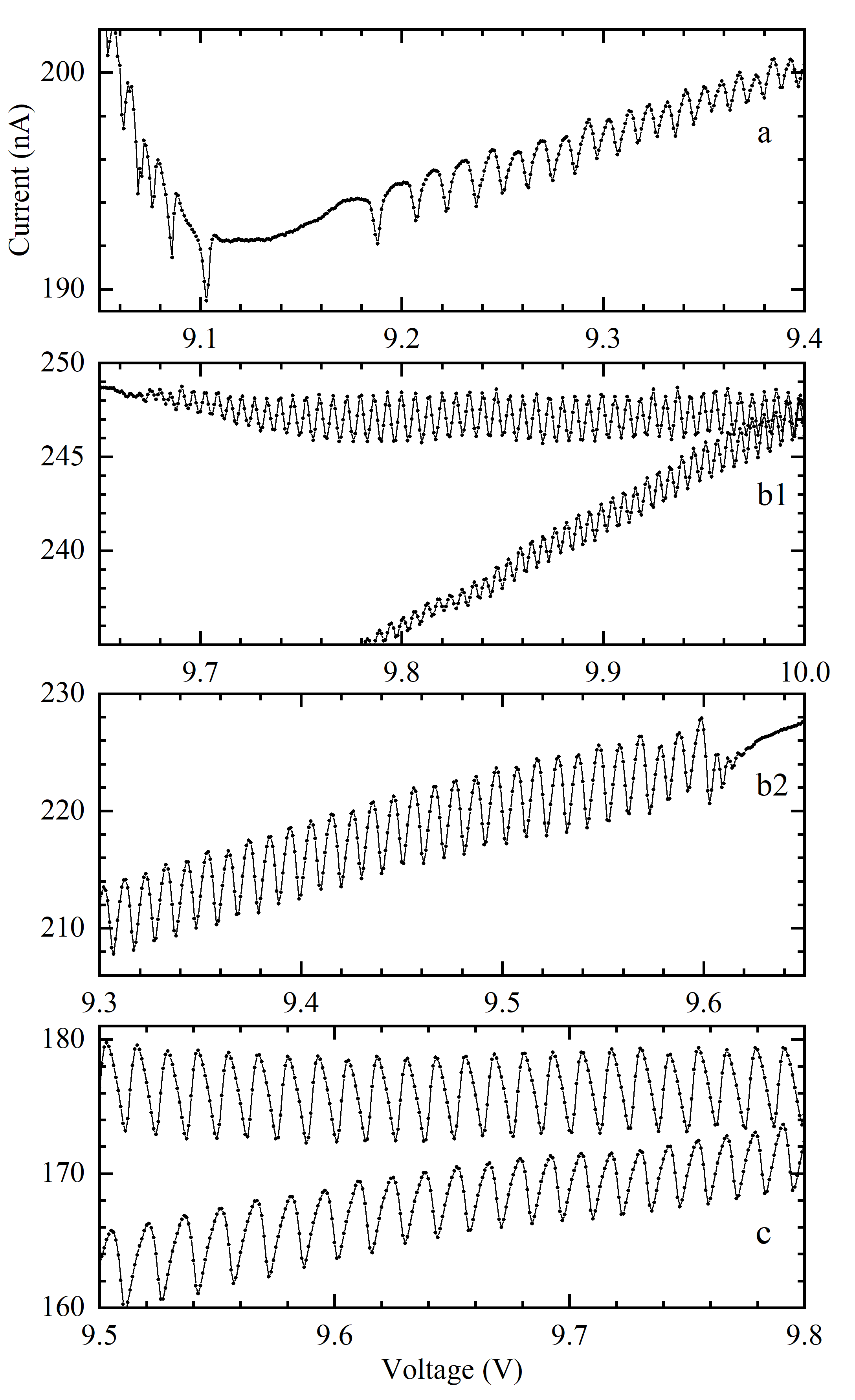}
    \caption{The corresponding panels in Fig.~\ref{fig:7} are enlarged. Apart from panel b1 which shows the developing oscillations near \SI{10}{\volt}, the panels show sharp defined “V-shaped” oscillations of similar amplitude. The period in panel a is seen to gradually reduce for increasing voltage.}
    \label{fig:8}
\end{figure}

A scan from \SIrange[range-phrase = --]{9}{10}{\volt} with a \SI{1}{\milli\volt} resolution in an alternative bending beam assembly is shown in Fig.~\ref{fig:9}. The line shape obtains a period of \SI{12}{\milli\volt} at \SI{10}{\volt}. An enlargement of the \SIrange[range-phrase = --]{9.75}{10}{\volt} range from panel a is shown in panel b, where the increasing voltage scan has been offset with \SI{-1}{\nano\ampere} for clarity. Similar to Fig.~\ref{fig:8} also this line has a sharp “V shaped” minimum. A spike of a single data point stands out on one of the flanks and reproduces over the stretch of the oscillations. The line-shape in the increasing voltage scan is continuing, current-axis mirrored, in the decreasing voltage scan inclusive of the \SIrange[range-units = brackets,range-phrase = --]{1}{2}{\milli\volt} features which are mirrored as well.

\begin{figure}[H]
    \centering
    \includegraphics[width=.49\textwidth]{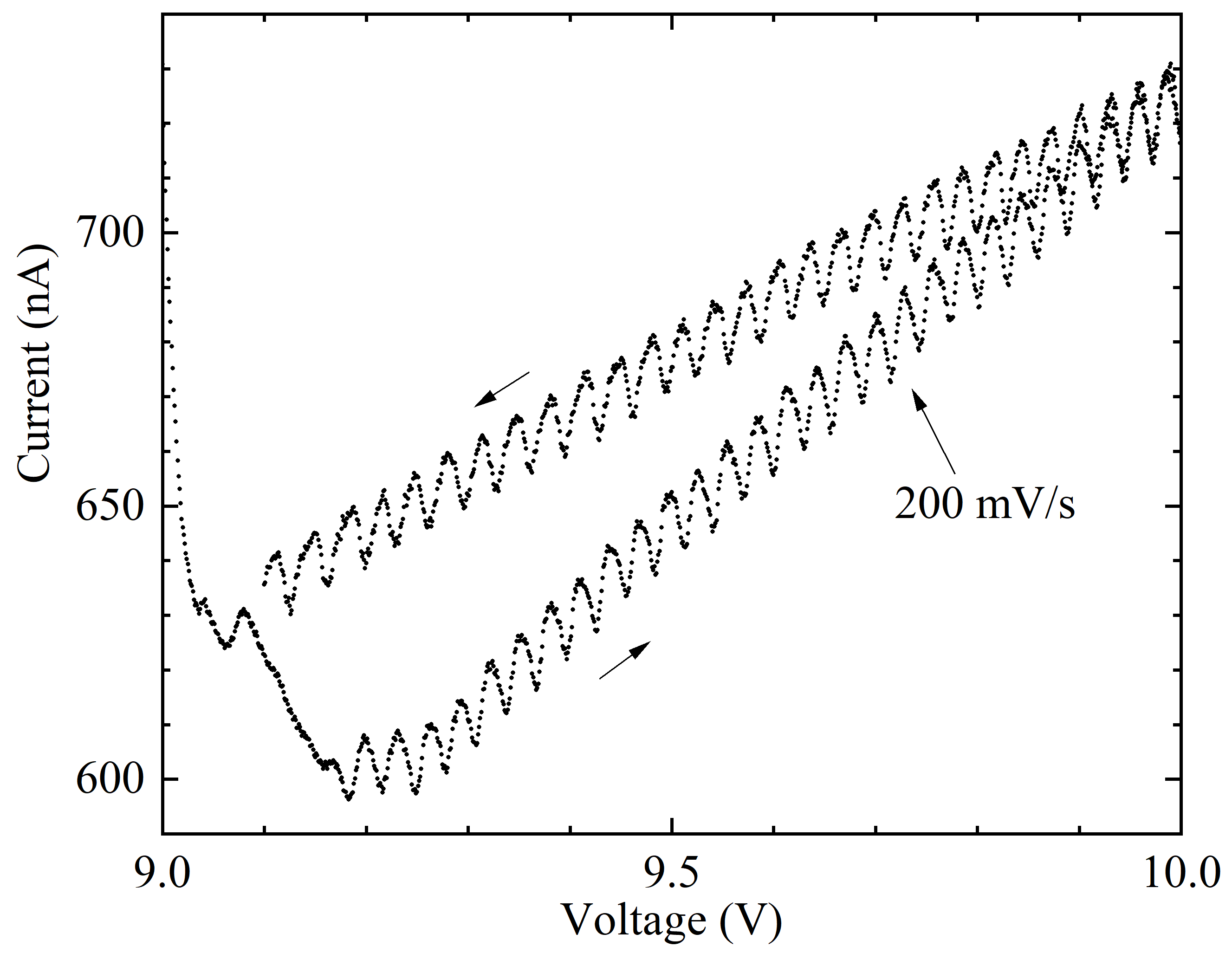}
    \caption{A deviating scan speed of \SI{1}{\volt} in 5 seconds has been used in the partially wet phase of this assembly to capture these "V-shaped" oscillations which extend over the entire voltage range, Au\textsuperscript{S}, $\phi_{\text{in}}^{2}$.}
    \label{fig:10}
\end{figure}

A scan which has been taken in 5 seconds in one direction is presented Fig.~\ref{fig:10}. The increasing and decreasing voltage scan have been recorded in one turn without interruption. A clear "switch on" effect is also present in this IV curve. The amplitude is of the same order as the oscillations in Fig.~\ref{fig:7} the period however is about \SI{30}{\milli\volt}, three times as large as compared to Fig.~\ref{fig:7}b.

\subsection{Oscillations in a BDT barrier THF partially-wet-phase junction}
The junctions in this section have all been immersed in a \SI{1}{\milli\Molar} BDT/THF solution. All results are obtained in the partially wet phase, after the draining and drying process. The goal is to capture an island in the form of a phenyl ring from the BDT molecule, coupled via two barriers to the opposing electrodes.

\begin{figure}[H]
    \centering
    \includegraphics[width=.49\textwidth]{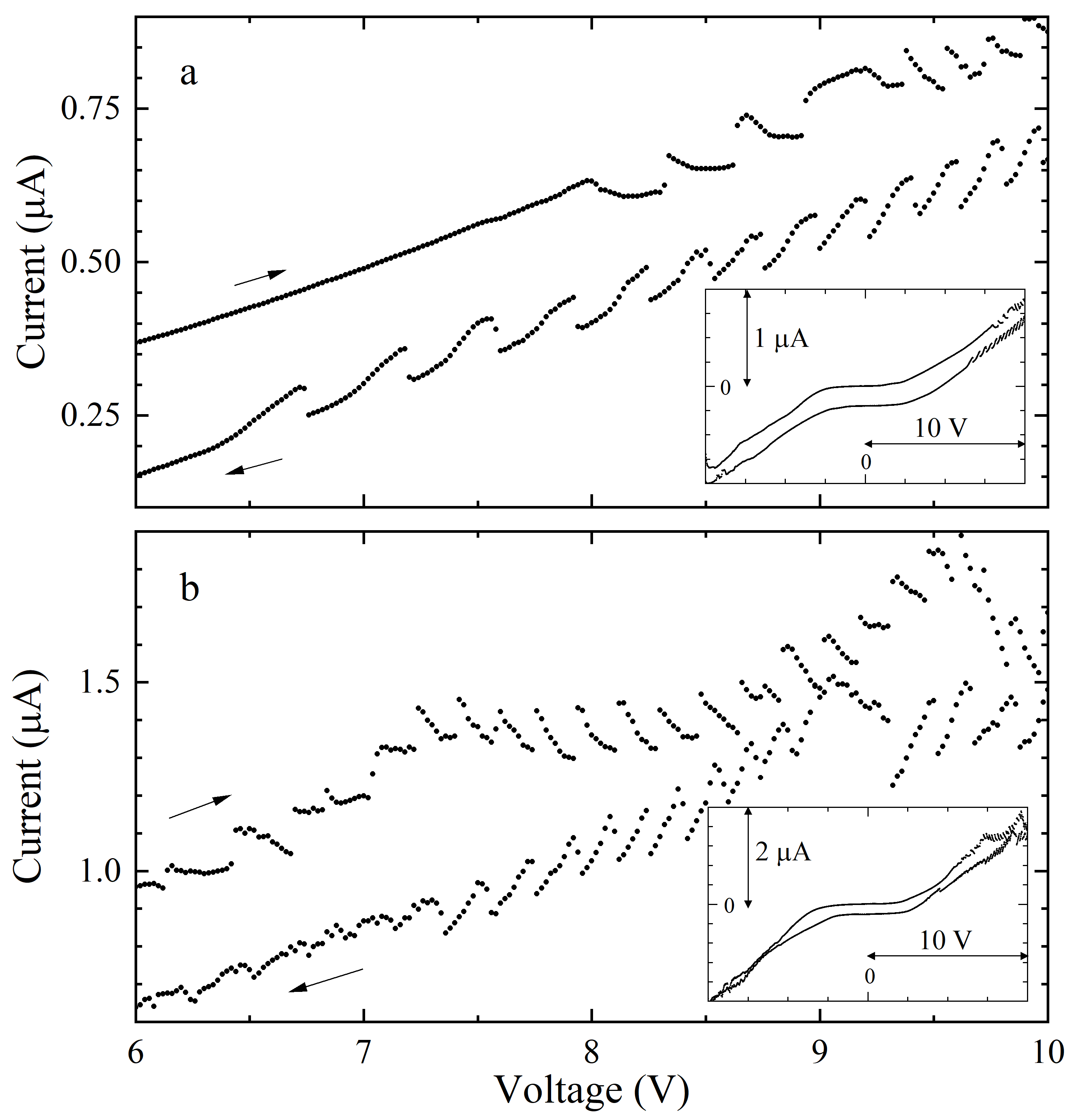}
    \caption{IV curves in the partially wet phase of two assemblies both exposed to the BDT/THF solution. Similarities in panel a and b are discontinuous current jumps trailed by a negative differential resistance in the increasing voltage scan and trailed by a positive differential resistance in the decreasing voltage scan. The insets show the increasing and decreasing scans over the entire \SIrange{-10}{10}{\volt} range. All decreasing voltage scans are offset by \SI{-200}{\nano\ampere} for clarity, panel a: Au\textsuperscript{H}, $\phi_{\text{in}}^{4.5}$, panel b: Au\textsuperscript{S}, $\phi_{\text{in}}^{4.5}$. }
    \label{fig:11}
\end{figure}

In Fig.~\ref{fig:11} an increasing and decreasing voltage trace is shown for two bending beam assemblies, one in panel \ref{fig:11}a the other in panel \ref{fig:11}b. The lower IV curves with a decreasing voltage are offset \SI{-0.2}{\micro\ampere} for clarity. The increasing and decreasing voltage scan are both part of the same scan. Clear discontinuous current steps are present in the observed structure. For the increasing scans a negative differential resistance section follows after each step, this section has some curvature before the next step occurs. For the decreasing scans the behavior is simply a continuation of the behavior in the increasing scan but mirrored in the current-axes. The curvature of the segments gets somewhat steeper towards higher voltages, which coincides with smaller voltage increments in subsequent steps. The insets show the IV curve over the full \SIrange{-10}{10}{\volt} range the lower curve is also offset with \SI{-0.2}{\micro\ampere}. The structure at positive bias does not reproduce at negative bias in both shown traces.

\begin{figure}[H]
    \centering
    \includegraphics[width=.49\textwidth]{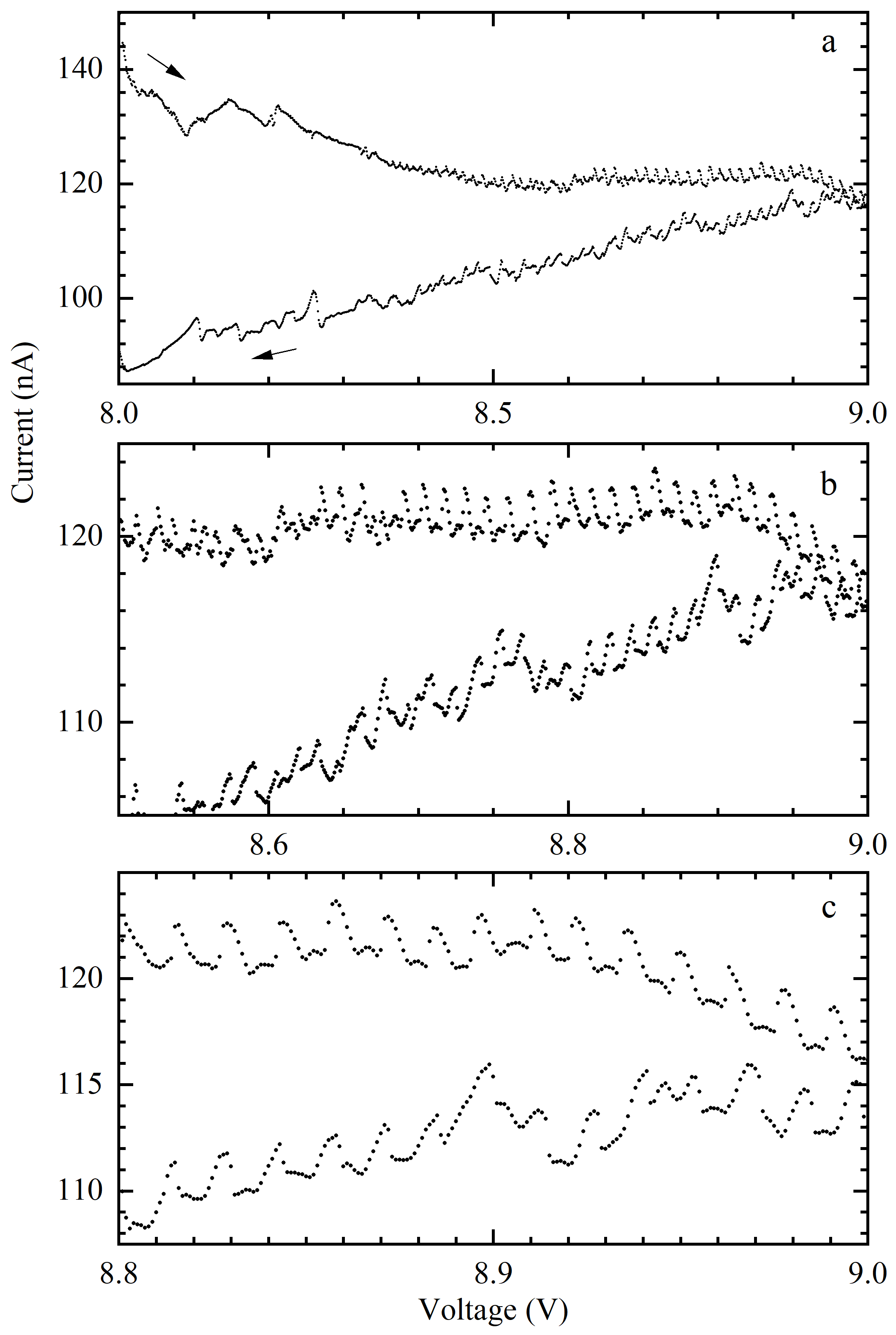}
    \caption{A high voltage trace of a BDT/THF junction in the partially wet phase, the panels a-c show an increasing enlargement. Similarities to the structure in Fig.~\ref{fig:11} are current discontinuities, trailing negative and positive differential resistance after a discontinuity depending on scan direction and current-axis mirrored structure upon scan reversal. In panel c the current of the lower curve is offset by \SI{-3}{\nano\ampere} for clarity, Au\textsuperscript{S}, $\phi_{\text{in}}^{2}$. }
    \label{fig:12}
\end{figure}

Also for the high voltage offset BDT barrier junctions the scans are not accompanied by the \SIrange{-10}{10}{\volt} scan. These scans are similar to the ones shown in the insets of Fig.~\ref{fig:6} and Fig.~\ref{fig:11}. Fig.~\ref{fig:12} presents data from a trace between \SI{8}{\volt} and \SI{9}{\volt}. Repetitive structure is present in both the increasing and the decreasing voltage scan. Panels b and c provide an enlargement, in panel c the lower IV curve has been offset \SI{-3}{\nano\ampere} for clarity. Over 40 oscillations in the upper curve and over 30 oscillations in the lower curve are counted. Similarly as in Fig.~\ref{fig:11} the shape of the oscillation for the decreasing voltage scan is current-axes mirrored of the shape in the increasing scan. The oscillation shape has also some resemblance to the line shapes shown in Fig.~\ref{fig:11}, there is a distinct discontinuous step present in every period. The periodicity of the oscillation changes from \SI{12}{\milli\volt} at \SI{8.4}{\volt} to \SI{14}{\milli\volt} at \SI{10}{\volt}.

\begin{figure}[H]
    \centering
    \includegraphics[width=.47\textwidth]{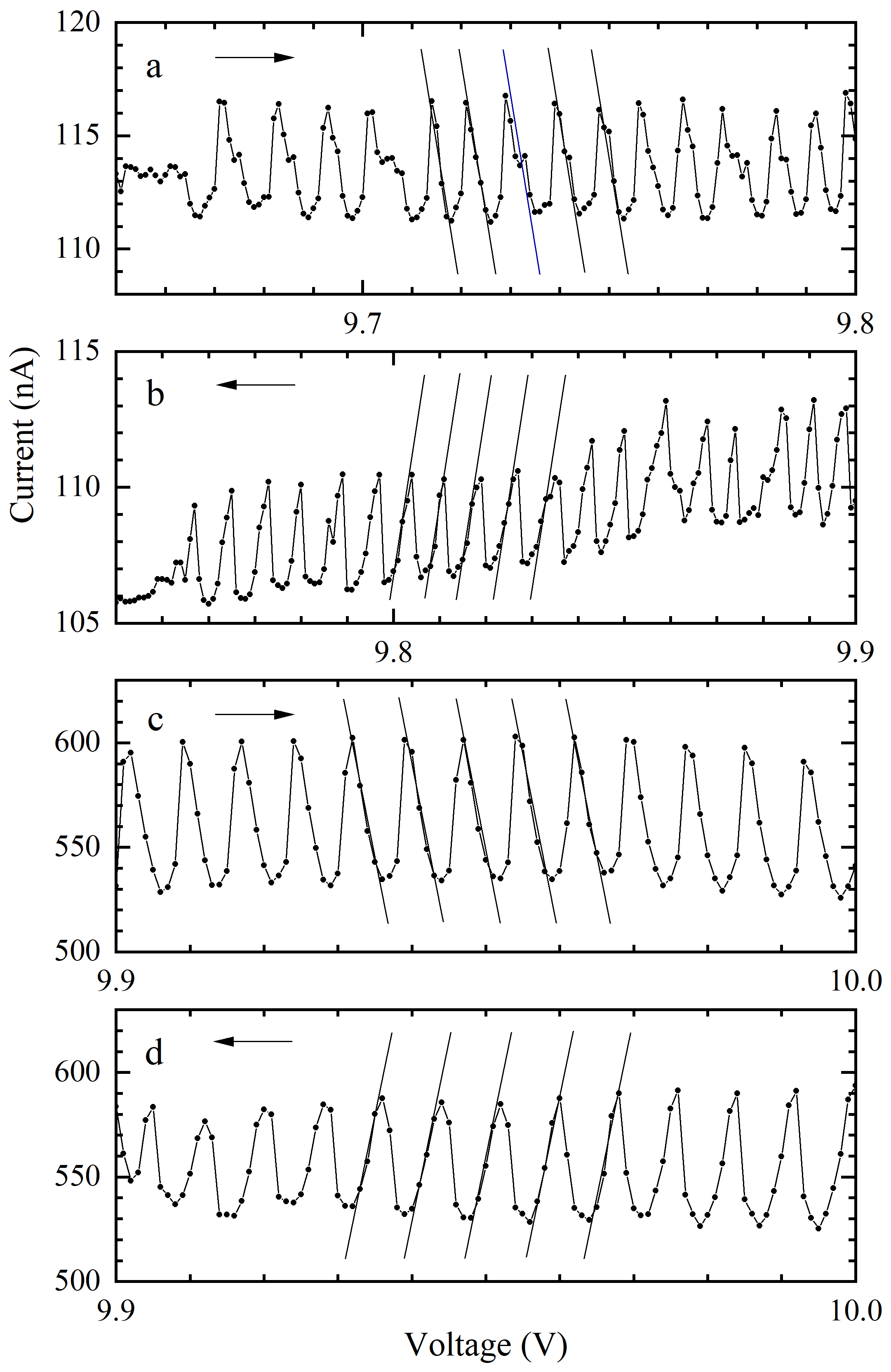}
    \caption{Panels a, b and c, d tie pair-wise to two bending beam assemblies. Increasing and decreasing scans replicate the main observations from Fig.~\ref{fig:12}. The lines indicating the slope are a guide to the eye. Panel a, b: Au\textsuperscript{S}, $\phi_{\text{in}}^{2}$, panel c, d: Au\textsuperscript{H}, $\phi_{\text{in}}^{4.5}$. }
    \label{fig:13}
\end{figure}

In Fig.~\ref{fig:13} data from 2 additional bending beam assemblies a/b and c/d show a similar result as indicated in Fig \ref{fig:12}. A regular pattern of peaks and discontinuous current jumps is clearly visible. Also in this case asymmetry in the peaks indicates a negative differential conductance trailing the current jump for the increasing scan and a positive differential conductance trailing the jump for the decreasing scan.

\begin{figure}[H]
    \centering
    \includegraphics[width=.46\textwidth]{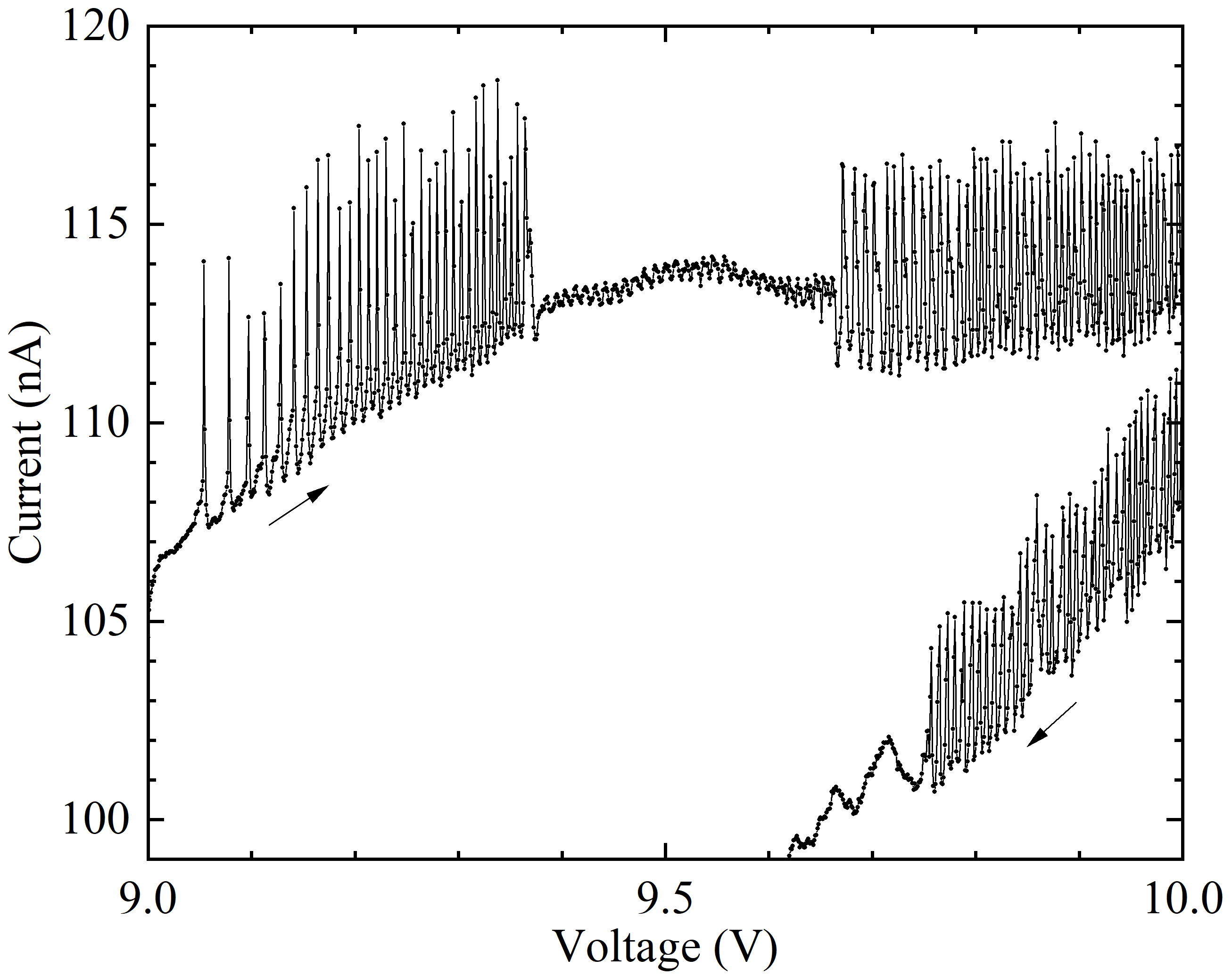}
    \caption{A high voltage scan over a \SI{1}{V} range, current discontinuities show as narrow spikes. The periodicity exceeding \SI{9.65}{V} is regular as was shown in Fig.~\ref{fig:13}a and b. The lower curve has been offset by \SI{-5}{\nano\ampere} for clarity. }
    \label{fig:14}
\end{figure}

The complete \SIrange[range-units = brackets,range-phrase = --]{9}{10}{\volt} scale of the data in Fig.~\ref{fig:13}a and b are presented in Fig.~\ref{fig:14}. Over 100 peaks are counted in this scan, the periodicity changes from \SI{24}{\milli\volt} near the start of the scan to \SI{6}{\milli\volt} at \SI{10}{\volt}.

\begin{figure}[H]
    \centering
    \includegraphics[width=.47\textwidth]{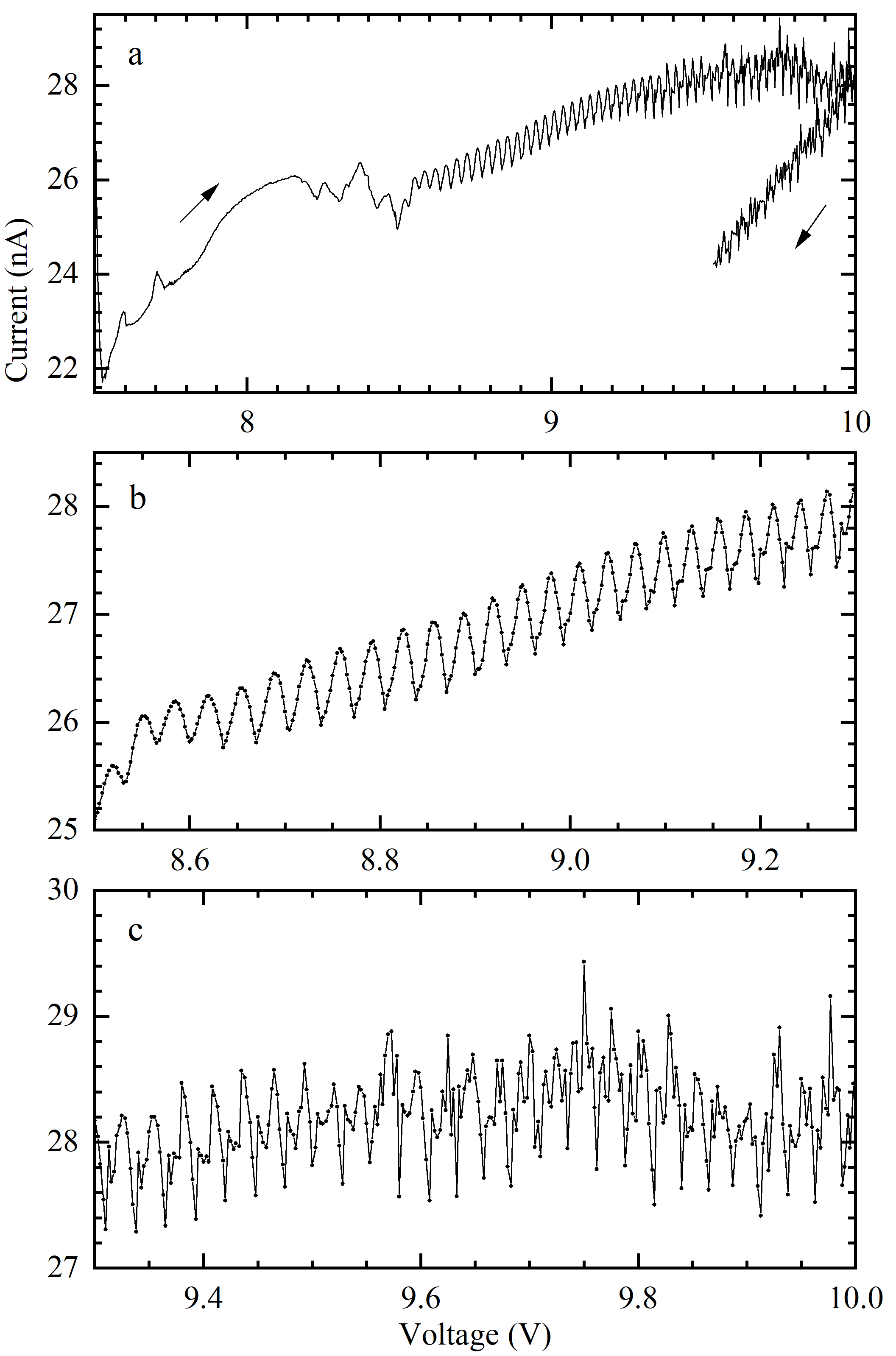}
    \caption{Although a BDT/THF solution has been used, the line shape resembles a THF barrier junction. Panels b and c show the increasing IV scan direction. The oscillations start of as “V-shaped” but quickly get more disorderly towards higher voltages, Au\textsuperscript{H}, $\phi_{\text{in}}^{4.5}$.}
    \label{fig:15}
\end{figure}

A markedly different line-shape is observed at the start of the oscillations in Fig.~\ref{fig:15}a with a scan extending from \SIrange{7.5}{10}{\volt}. The initial line shape exhibits a very sharp “V” shaped minimum and a somewhat rounded top as observed in the previous section. For increasing voltages exceeding \SI{8.75}{\volt} a “shoulder” is developing which is shifting with respect to the sharp bottom of the line as shown in Fig.~\ref{fig:15}b. Fig.~\ref{fig:15}c shows the development of a smooth line-shape into a more chaotic one, although the repetition of the sharp minima reflects the presence of the periodicity all the way to \SI{10}{\volt}.

\begin{figure}[H]
    \centering
    \includegraphics[width=.45\textwidth]{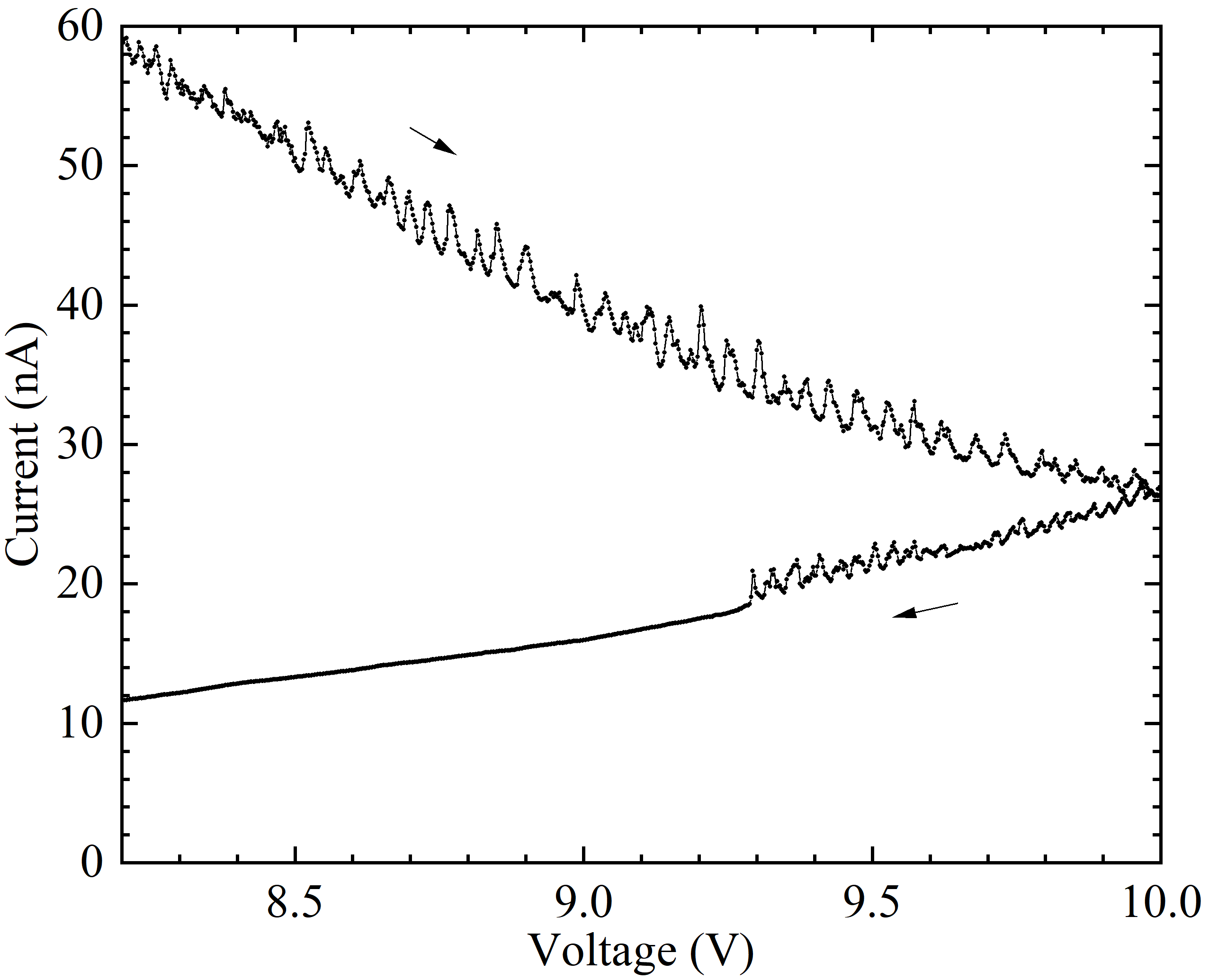}
    \caption{The partially wet phase abruptly ends in the decreasing voltage scan at \SI{9.3}{\volt}. The irregular oscillations show typical BDT barrier behavior. Once the partially wet phase seizes to exists a smooth structure-less IV curve is the result, Au\textsuperscript{H}, $\phi_{\text{in}}^{4.5}$.}
    \label{fig:16}
\end{figure}

The transition to the completely dry phase is indicated in Fig.~\ref{fig:16}. A steep decrease in conduction, with a lot of structure, is abruptly transformed in a line without any structure. This is the point where the partially wet phase is expected to transit abruptly to the completely dry phase as for example also noted in the inset of Fig.~\ref{fig:6}a.

\subsection{Large voltage range molecular spectra}
In the previous sections it has been shown that some of the high voltage fine structure of the overall line-shape for THF barrier and BDT barrier junctions in the THF partially wet phase reproduce. Is it possible to also reproduce partially wet phase molecular spectra over a large voltage range?

\begin{figure}[H]
    \centering
    \includegraphics[width=.49\textwidth]{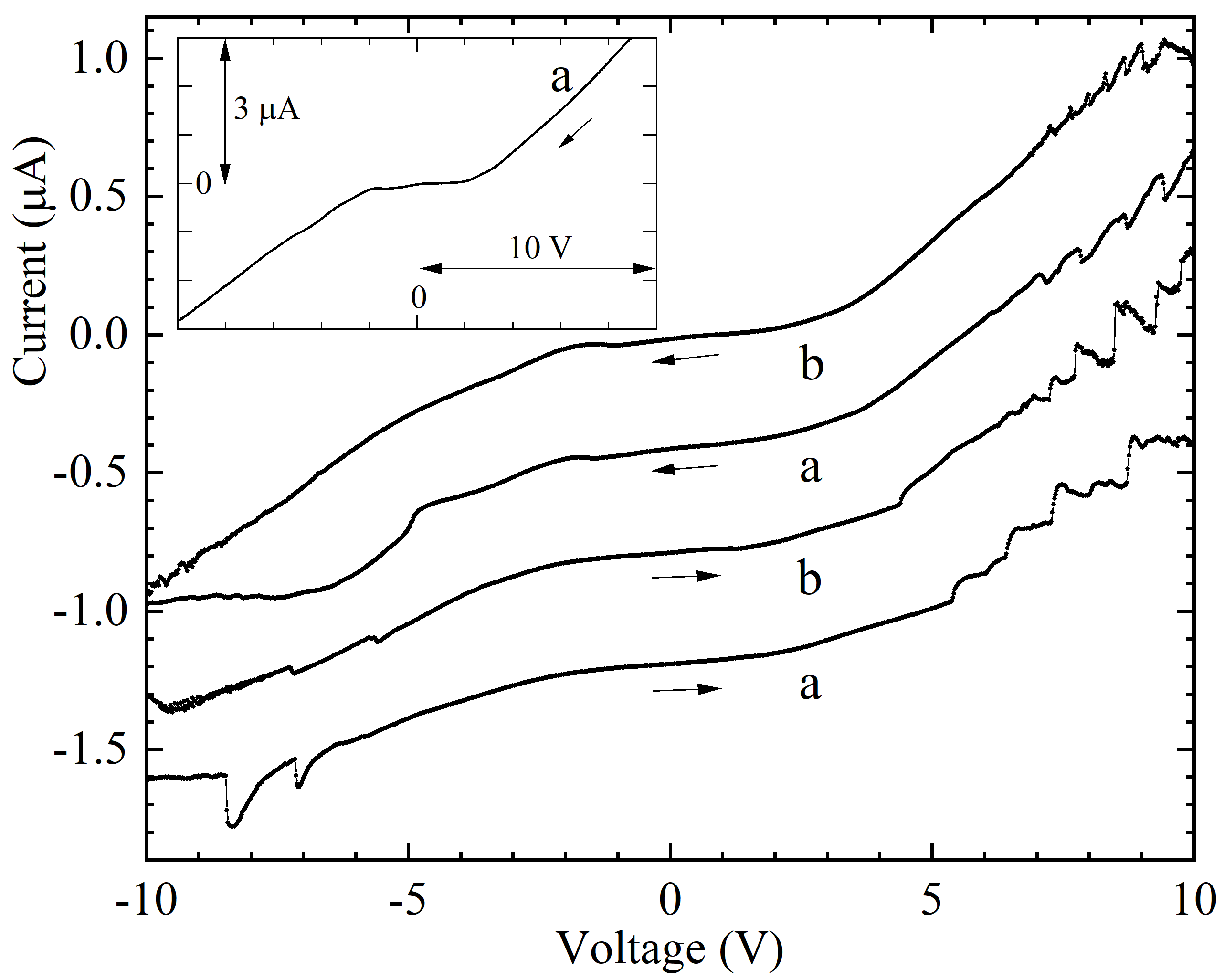}
    \caption{Large voltage scale molecular IV spectra of bending beam assembly a and b. Curves have been offset by \SI{-0.4}{\micro\ampere} for clarity. The inset shows how IV curve a has progressed after 15 minutes, just prior to entering the completely dry phase, a,b: Au\textsuperscript{S}, $\phi_{\text{in}}^{4.5}$.}
    \label{fig:17}
\end{figure}

Devices in Fig.~\ref{fig:17} have been immersed in the BDT/THF solution in the $\phi_{\text{in}}^{4.5}$ cell. Two consecutive voltage scans for bending beam assemblies a and b in the partially wet phase are presented. The IV curves are offset by \SI{0.4}{\micro\ampere} with respect to each other. For the increasing scans two discontinuities are visible at negative bias for which the amplitude and exact positions differs. The observed structure at positive bias shows some resemblance to what was shown in Fig.~\ref{fig:11}. In this case there are plateaus, there is structure on the plateaus, there are discontinuous steps as well but the repetition is more irregular and less in number as compared to the structure in Fig.~\ref{fig:11}. For the decreasing voltage scans the structure at positive bias resembles that in Fig.~\ref{fig:11} in that the discontinuous steps are trailed by a positive differential resistance, however the repetition is more irregular and there are far less steps. For the scans in negative direction there is an inflexion point at approximately \SI{-2}{\volt} and \SI{-5}{\volt}. The inset shows how the IV curve from assembly a has progressed over time after approximately 15 minutes from recording the IV curves in the main figure. The structure has disappeared and the current extends to \SI{3}{\micro\ampere} at \SI{10}{\volt}.

\begin{figure}[H]
    \centering
    \includegraphics[width=.49\textwidth]{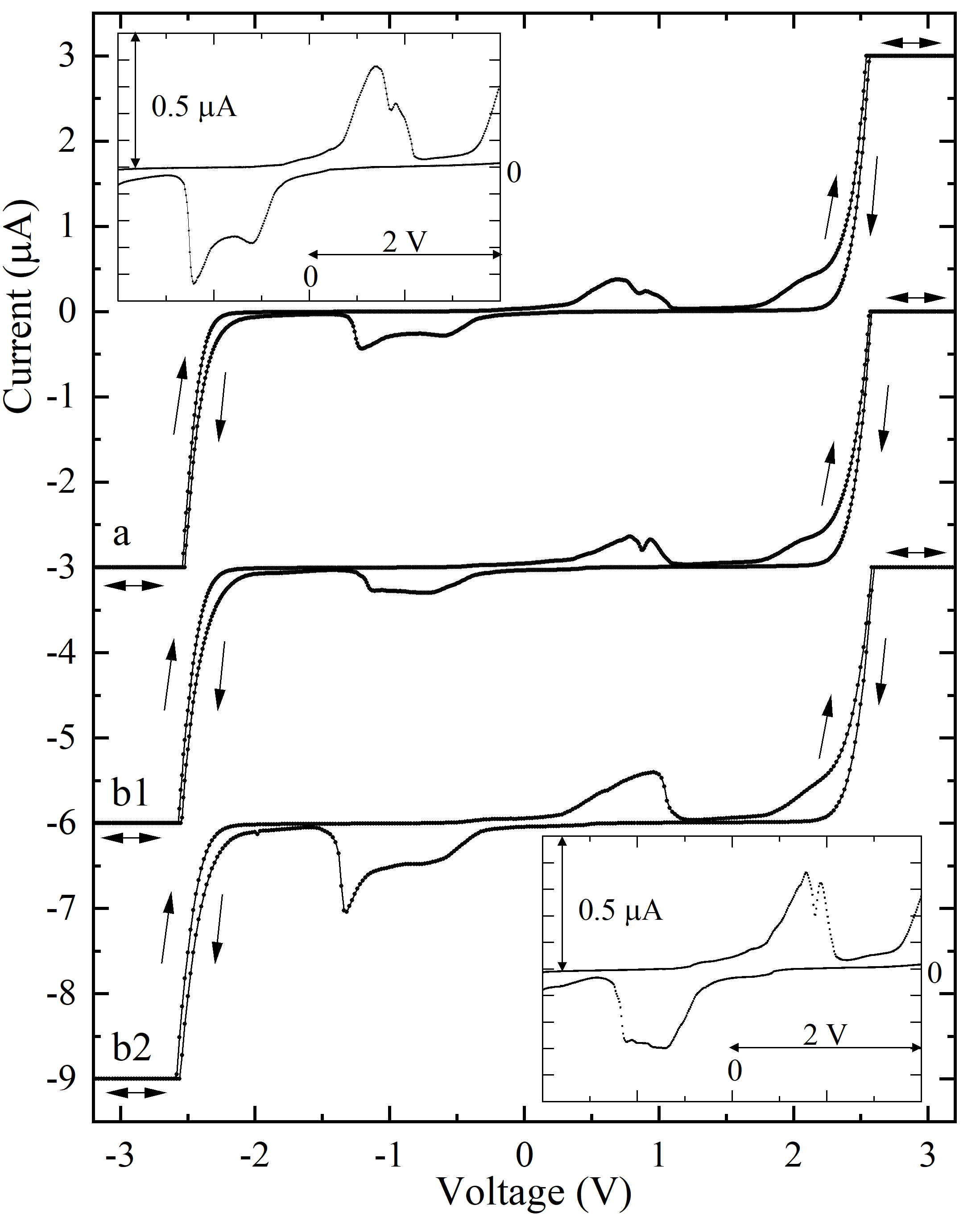}
    \caption{IV curves in the H$_2$O partially wet phase. Compliance was set at $\pm\SI{3}{\micro\ampere}$, the curves are offset with \SI{3}{\micro\ampere} for clarity. The IV curves are very similar and are stable over hours. The sub-gap current is dependent on the voltage scan speed as indicated by curve b2 which was obtained with a double scan-speed as compared to b1. The insets provide an enlargement of the central section for curves a and b1, a,b: Au\textsuperscript{S}, $\phi_{\text{in}}^{2}$.}
    \label{fig:18}
\end{figure}

IV traces of two bending beam assemblies, a and b, with an H$_2$O partial wet phase are shown in Fig.~\ref{fig:18}. These junctions are created by starting off with a bending beam assembly immersed in a BDT/THF solution. The procedure in order to arrive at a molecular junction in the THF partially wet phase as described above has been followed to the point where during the formation of the junction the ambient H$_2$O humidity was increased to \SI{95}{\percent} by using a humidifier within a contained environment. The \SI{95}{\percent} humidity level was reached in a matter of seconds, the humidifier was switched off and the system was allowed to gradually transit to an ambient humidity level of \SI{60}{\percent} in a few hours’ time. A gradual transition from the default BDT IV curve (see for example insets Fig.~\ref{fig:6}, Fig.~\ref{fig:11}) to the IV curves shown in Fig.~\ref{fig:18} took place. The IV curves have been recorded in a continuous scan mode between \SI{-5}{\volt} and \SI{5}{\volt}, the current level compliance was set at \SI{-3}{\micro\ampere} and \SI{3}{\micro\ampere}. Once reaching their final shape the IV curves a and b1 were stable for hours. Doubling the voltage scan speed, see curve b2, left the IV curve as is however the sub-gap structure near \SI{-1}{\volt} and \SI{1}{\volt}, approximately doubled in height. The insets show an enlargement of the central section of curves a and b1. Fig.~\ref{fig:18} shows the only data concerning H$_2$O partially wet phase junctions.

\section{Theoretical Background and Discussion}

\subsection{One-dimensional conduction}
\begin{figure}[H]
    \centering
    \includegraphics[width=.45\textwidth]{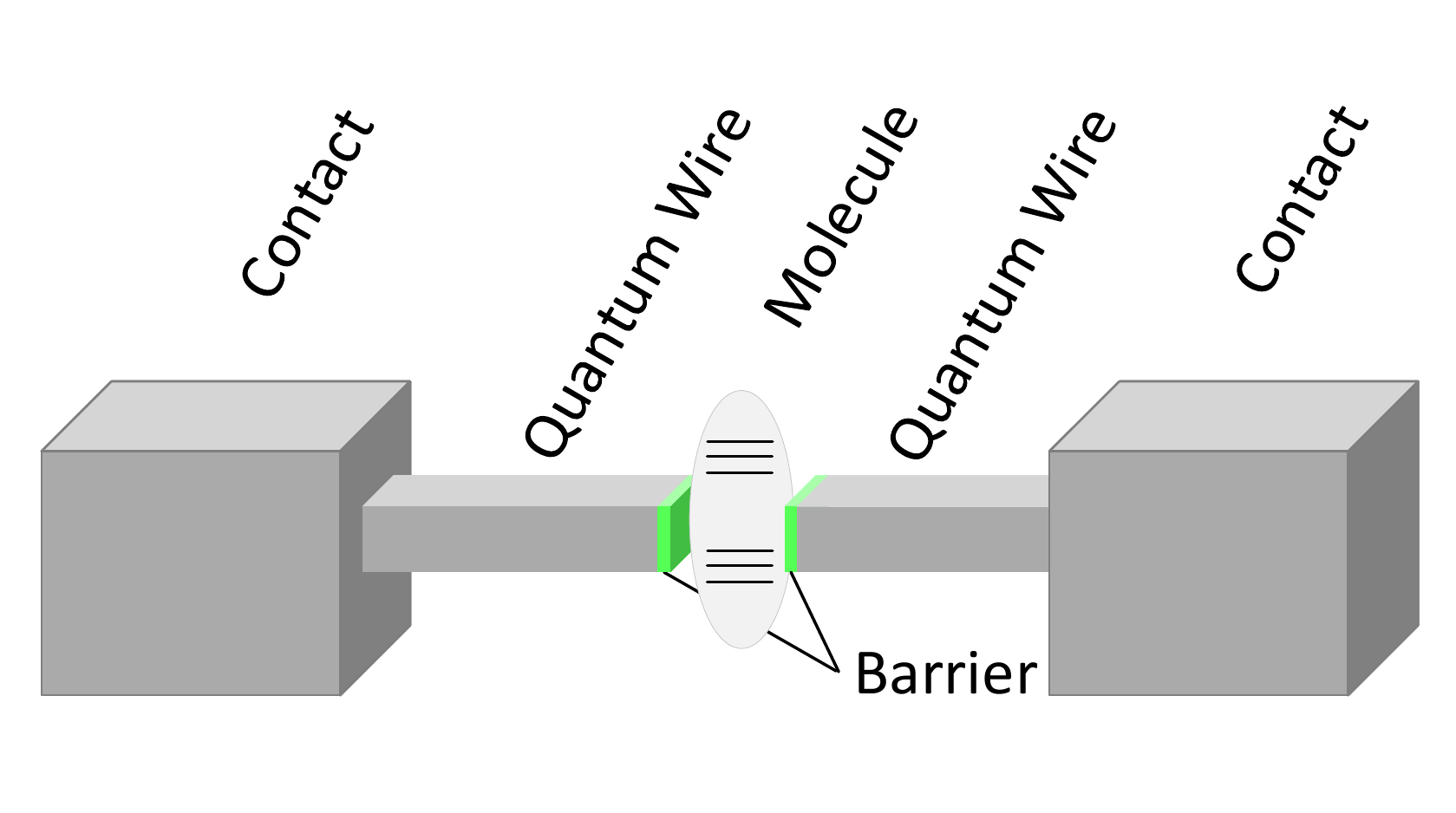}
    \caption{A schematic representation of the junctions under study. The molecule is connected via two barriers to a quantum wire on either side.  }
    \label{fig:19}
\end{figure}

From the data in section 3 it is clear that the conductance of the devices under study are of the order of, or well below \num{e-3} $2e^2/h$. This indicates that the physics is described by transmission and reflection, deep in the Landauer regime \mcite{landauer1957one,*landauer1981two,*landauer1989three}. Fig.~\ref{fig:19} provides a schematic representation of the junctions which are studied. 

\begin{figure}[H]
    \centering
    \includegraphics[width=.49\textwidth]{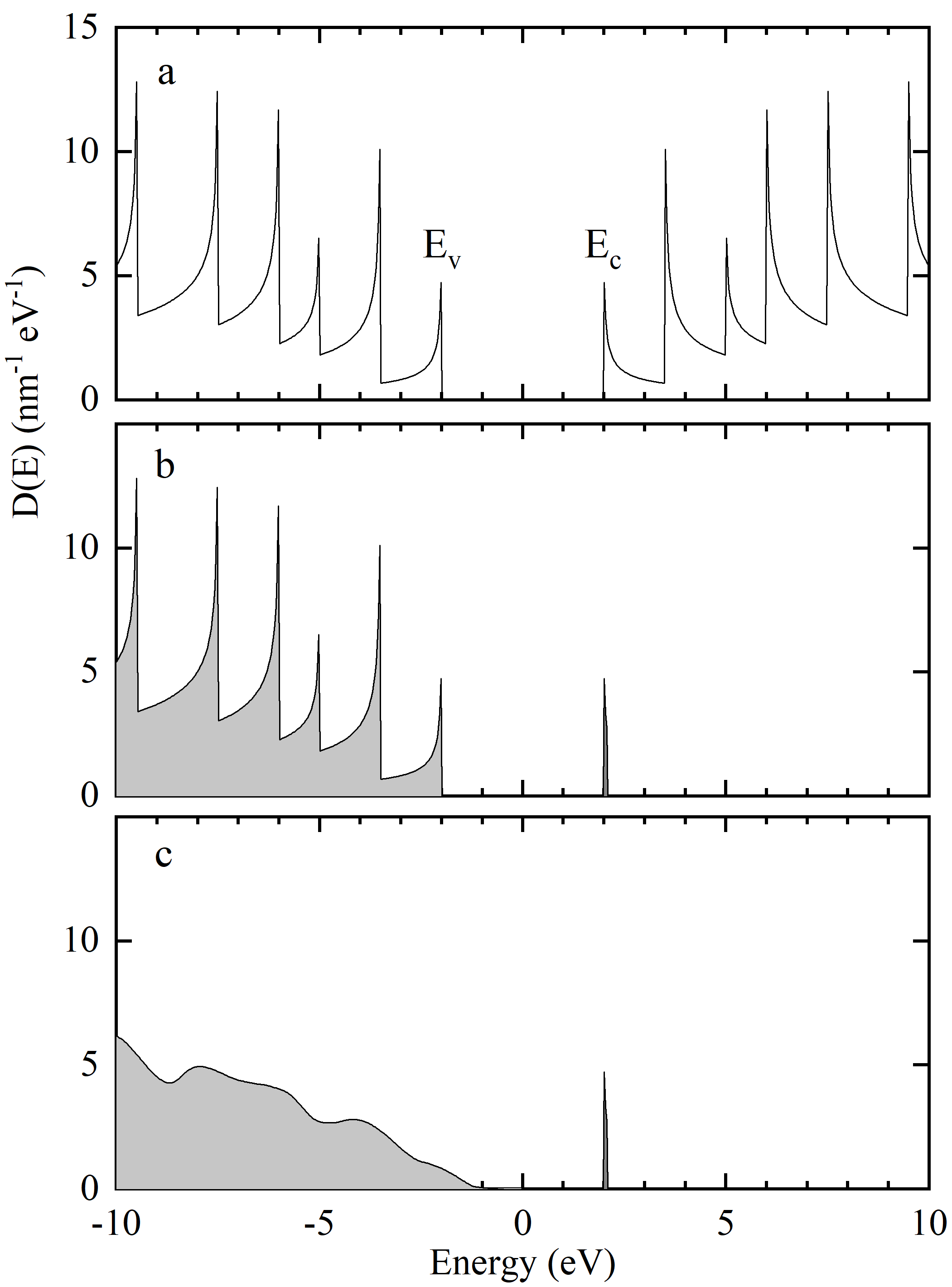}
    \caption{a. One dimensional density of states for a free electron gas, estimated input-parameters: $L_z=L_y$ equals \SI{8.6}{\angstrom} leads to $\varepsilon_y=\varepsilon_z=\SI{0.5}{\electronvolt}$, $E_c- E_v=\SI{4}{\electronvolt}$. Panel b shows the filling of the modes up to and including the lowest conduction sub-band. In panel c the valence sub-bands are allowed to interact and couple together. }
    \label{fig:20}
\end{figure}

The central molecule is connected via tunnel barriers to the two quantum wires, extending in the x-direction with a length $L_x$, a width $L_z$ and a height $L_y$. For a free electron gas the energy of the conduction modes in the quantum wires is described by: $E_{n,p}=E_c+n^2 \varepsilon_y + p^2 \varepsilon_z + (\hbar k_x)^2/2m$, likewise the relation of the valence modes is given by: $E_{n,p}=E_v-n^2 \varepsilon_y - p^2 \varepsilon_z - (\hbar k_x)^2/2m$, where $E_c-E_v$ is the gap between the lowest conduction band and the highest valence band, $m$ is the electron mass and $\varepsilon_y$, $\varepsilon_z$ are provided by: $(\pi \hbar)^2/2mL^2_y$, $(\pi \hbar)^2/2mL^2_z$ respectively, $n$ , $p$ are integers $>0$. Neglecting spin-degeneracy and taking $L_z=L_y$ equals \SI{8.6}{\angstrom} or 3 Au atom diameters leads to \SI{0.5}{\electronvolt} for $\varepsilon_y$, $\varepsilon_z$. The density of states of the quantum wires is provided in Fig.~\ref{fig:20}a where the estimate for $E_c- E_v$ is \SI{4}{\electronvolt}, $n$, $p$ runs from 1 to 4 and all modes are counted. It is noted that the inputted numbers are rough estimates to be able to provide a qualitative picture of the conduction through a single molecule. In Fig.~\ref{fig:20}b only one conduction mode is populated, the lowest $n=1$, $p=1$ mode is responsible for the conduction in this case the higher energy conduction modes are empty. Finally in Fig.~\ref{fig:20}c the filled valence modes are allowed to interact and couple, leading to a smoother density of states as indicated qualitatively. Derived from first principles Grigoriev et al. \cite{grigoriev2006critical} arrived at a gap pinned to the Fermi level for a BDT molecule bridged between two Au electrodes, which supports the described experimental observations.

For the voltage biased systems under study it is notable that almost the entire voltage will be biased across the molecule and associated barriers. The voltage over the quantum wire typically amounts to less than one or a few \si{\milli\volt} even if the molecule is biased at \SI{10}{\volt}. This is due to the large difference in the total conduction with respect to $2e^2/h$ of conduction per mode in the quantum wire. The density of states of the quantum wire in Fig.~\ref{fig:20}c is voltage independent and will simply shift “up and down” with an increasing or decreasing voltage bias over the contacts, even if this amounts to \SI{10}{\volt}. Is \SI{10}{\volt} not an extremely high voltage to bias a single molecule with? Molecules exhibit efficient tunnel barriers, capable of sustaining high \si{\electronvolt} molecular energy levels.  In addition, a single molecule can consist of multiple barriers \cite{grigoriev2006critical}, for example a bridging BDT molecule exhibits 4 tunnel barriers reducing the bias per barrier:
\begin{itemize}
    \item Barrier 1 from the 1\textsuperscript{st} Au electrode to the 1st S atom
    \item Barrier 2 from the 1\textsuperscript{st} S atom to the phenyl ring
    \item Barrier 3 from the phenyl ring to the 2\textsuperscript{nd} S atom
    \item Barrier 4 from the 2\textsuperscript{nd} S atom the 2\textsuperscript{nd} Au electrode
\end{itemize}
Finally, the experiment allows and caters for this. Dissipation will in accordance with the Landauer theory take place far away from the molecule and quantum wires, in the electrode banks where energetic electrons will be able to get into thermal equilibrium with the lattice. 

What is the experimental evidence in support of one dimensional conduction? To answer this question the process of how to arrive at partially wet phase junctions is detailed. The process is split into two: a formation process followed by a nucleation process.  During the formation process excess fluid is slowly evaporating, BDT molecules are attaching to surfaces and the two electrodes are slowly getting ever closer together, continuously increasing the conductivity. The conduction still increases over time for a THF partially wet phase BDT junction, after the formation process has stopped. This increase is attributed to physi-sorption of one or more molecules to the bridging molecule. The increasing size of the central island leads to a denser energy level spacing thus increasing conductivity while still maintaining a single point of tunneling related to both electrodes. This process continues until the nucleated island becomes too big and starts to touch one of the two electrodes, creating multiple conduction points and thus losing the one dimensionality. 

The data in Fig.~\ref{fig:17} have been carefully selected to be similar at a specific point in the nucleation process from a data set of 20 different bending beam assemblies. The data exhibit structure in the IV curve which disappears altogether into a higher conductivity curve by waiting 15 minutes see the inset in Fig.~\ref{fig:17}. Even though the a and b curves are not identical, they do show similar features for the increasing as well as the decreasing curves. In general the $\phi_{\text{in}}^{4.5}$ cell shows more nucleation as compared to the $\phi_{\text{in}}^{2}$ cell as defined by the conductivity of the final IV curve just prior to entering the completely dry phase. The $\phi_{\text{in}}^{4.5}$ cell can conduct \SI{3}{\micro\ampere} at \SI{10}{\volt} for a BDT barrier junction, for the $\phi_{\text{in}}^{2}$ cell a \SI{1}{\micro\ampere} level is the norm. This is attributed to an easier access of the junction to air molecules in the $\phi_{\text{in}}^{4.5}$ cell.

The data in Fig.~\ref{fig:18} stand out as halfway the BDT THF formation process the H$_2$O humidity was increased to \SI{95}{\percent}. It is astonishing that this system reorganizes itself to a stable molecular junction for hours with an H$_2$O partial wet phase layer. The collapse into a completely dry phase is not observed, nor is the nucleation phase observed. Apparently the H$_2$O partially wet phase layer protects the bridging BDT molecule from other molecules nucleating on it. The fact that the H$_2$O partially wet phase remains intact indefinitely at ambient conditions is to be expected \cite{bonn2009wetting}. These data have not been selected. The same reproducible enduring end-state IV curves indicated in Fig.~\ref{fig:18} result from this experiment in 4 out of 8 bending beam assemblies.

The sub-gap current peak in the curves in Fig.~\ref{fig:18} shows voltammetry behavior: the current peak at approximately 1V in the positive direction reproduces in the opposing scan direction at reverse polarity. What is the source of this current? As the electrodes are not in the wet phase there can be no chemical electrode surface reaction responsible. The dipole moment of the H$_2$O molecule is expected to play a determining role. At a certain voltage bias the dipole-moment of all the partially wet phase H$_2$O molecules on the biased electrode will direct themselves with their electropositive or electronegative side towards the electrode, depending on the polarity of the bias. These aligned dipoles will hold some opposite charge, residing on the electrodes, captive. Once the dipole is flipped by a changing polarity on the electrode all the charge that was held captive becomes available and creates a displacement current proportional to $\dv*{V}{t}$. From Fig.~\ref{fig:18} it is obvious that the flipping of the dipoles occurs in a range around \SI{-1}{\volt} and \SI{1}{\volt}, releasing about \SI{e-6}{\coulomb} as obtained from integrating the sub-gap current peak over the time it was measured in. Doubling the voltage scan rate in curve 2b in Fig.~\ref{fig:18} also approximately doubles the measured sub-gap current which is consistent with an independent constant total captive charge, released upon flipping of the dipoles. Partially wet phase voltammetry can be used as a means of detection of H$_2$O in for example the THF partially wet phase layer, by studying the presence of structure at \SI{-1}{\volt} and at \SI{1}{\volt}.

The two main experimental findings in support of 1D conduction are:
\begin{enumerate}
    \item Obtaining the same IV curves for multiple BDT barrier H$_2$O partially wet phase junctions is a strong indication that the same singular molecular junction is created. This is indicative that also the THF barrier and BDT barrier THF partially wet phase junctions are of singular nature.
    \item The regular occurrence of negative differential resistance in the IV curves in section 3.2 and 3.3 also is a strong indication of 1D conduction. 
\end{enumerate}

\subsection{Grundlach oscillations}
\begin{figure}[H]
    \centering
    \includegraphics[width=.3\textwidth]{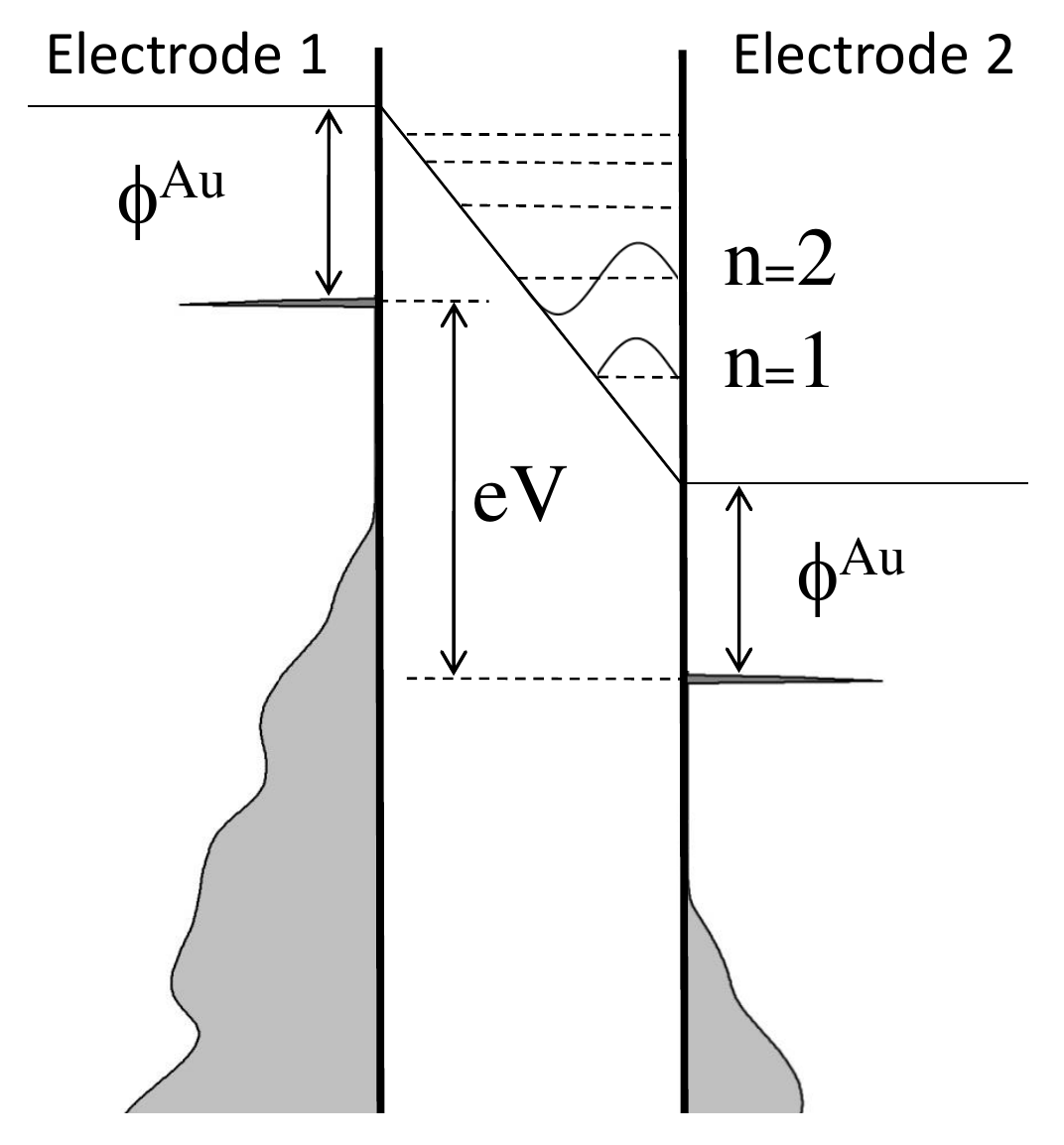}
    \caption{Explanation of the Grundlach resonances. Standing waves are able to build in the trapezoidal resonator. The one dimensional density of states peak at the Fermi level is ideally suited to probe these states, explaining the observed current oscillations in IV curves.}
    \label{fig:21}
\end{figure}

Oscillations in the conductance as a function of energy in a planar metal tunnel junction biased with a voltage larger than the work functions of the electrodes were predicted by Grundlach in 1966 \cite{gundlach1966berechnung}. Grundlach oscillations, or also field emission resonances, can be viewed as electron standing waves in the resonator bounded by the barrier created by the field on one side and the surface potential step on the other side see Fig.~\ref{fig:21}. Gundlach oscillations have been observed in planar junctions \cite{GLplanar1974tunneling}, in high vacuum STM junctions \cite{binnig1982helv} as well as in MCB junctions at cryogenic temperatures \cite{kolesnychenko2000field}. Solving the 1 dimensional Schr\"odinger equation with a trapezoidal barrier and a surface potential step at electrode 2 leads to standing waves at specific energy eigenvalues $E_n$ in the well:

\begin{equation}\label{eq:1}
    E_n=\left(\frac{3\pi\hbar e}{2\sqrt{2m}} \right)^{2/3} \varepsilon^{2/3}\left(n-\frac{1}{4}\right)^{2/3}\;	n=1,2,...   
\end{equation}

An effective electric field inside the barrier is represented by $\varepsilon$. At the surface potential step in electrode 2 there is no barrier, the energy spectrum is continuous and electrons can enter electrode 2 at any energy above $\Phi^{\text{Au}}$. Fig.~\ref{fig:21} shows the junction conductance to peak whenever the bias voltage reaches one of the levels $E_n$.

\begin{equation}\label{eq:2}
   eV = \Phi^{\text{Au}}+\left(\frac{3\pi\hbar e}{2\sqrt{2m}} \right)^{2/3} \varepsilon^{2/3}\left(n-\frac{1}{4}\right)^{2/3}
\end{equation}

In Fig.~\ref{fig:21} the one dimensional density of states from Fig.~\ref{fig:20}c is included for electrode 1 and 2. For the vast majority of the data obtained from partially wet phase junctions, created in the way described above, current oscillations are detected. This is consistent with a one dimensional density of states. With an increasing bias the oscillations in the current in Fig.~\ref{fig:5} for example, are explained in a natural way. Whenever the peak at the Fermi level coincides with an energy state the current is maximized, if the Fermi level is in between states the current is reduced. At some bias greater than $E_c-E_v$ the valence band states starts to gradually contribute to the current as well.

\begin{figure}[H]
    \centering
    \includegraphics[width=.49\textwidth]{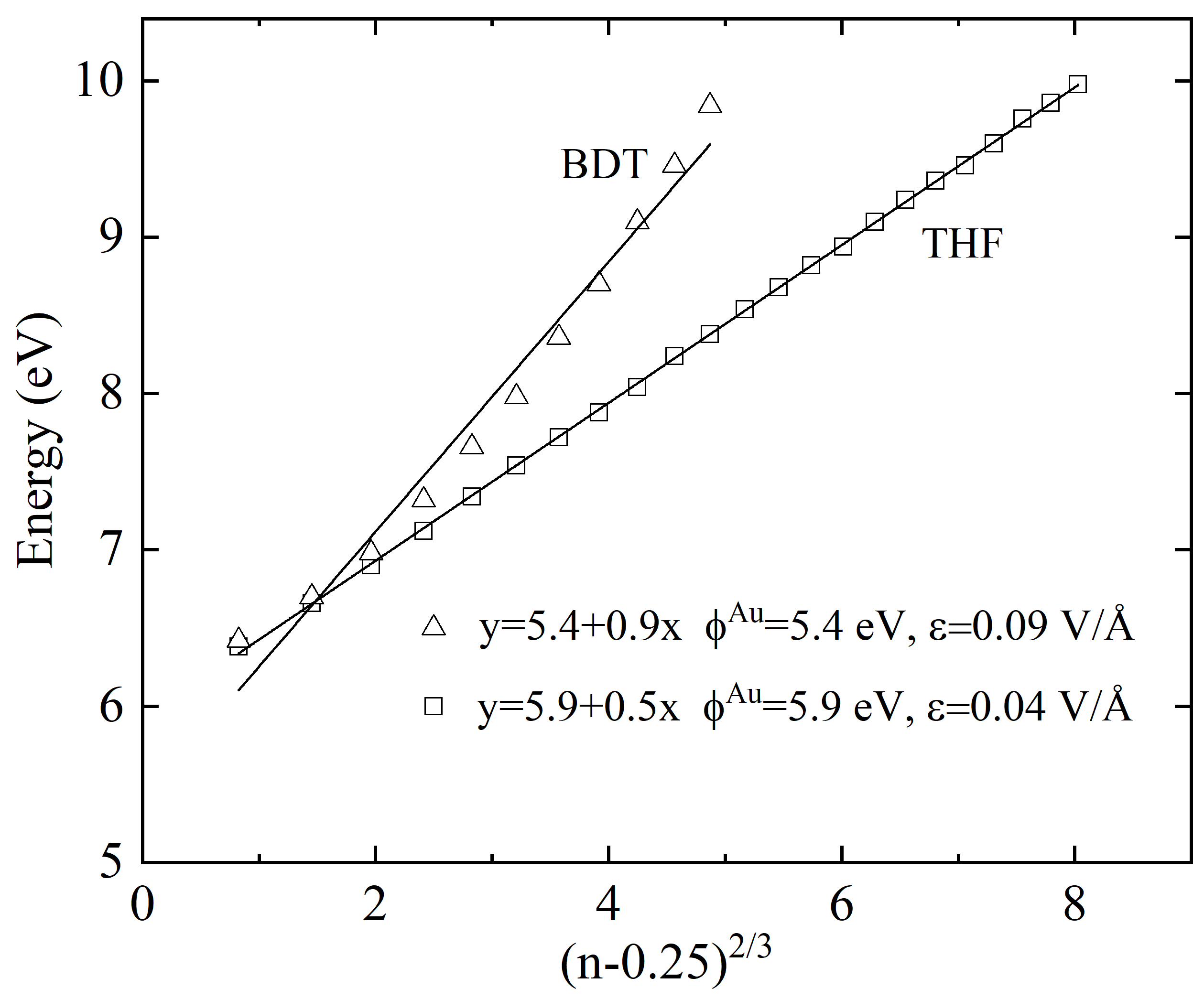}
    \caption{Energy of the observed peak in the IV curve versus peak index revealing the work-function $\Phi^\text{Au}$ as well as the effective electric field $\varepsilon$}
    \label{fig:22}
\end{figure}

In Fig.~\ref{fig:22} the peak number n and peak energy values are displayed for data from Fig.~\ref{fig:4}c and Fig.~\ref{fig:5}. The work-function from gold $\Phi^{\text{Au}}$ as well as the effective electric field $\varepsilon$ can be directly determined from the interception with the energy axis and the slope of the graph respectively. Values are indicated in the graph, the values for $\varepsilon$ are low, work-function values are in line with what is reported in \cite{kolesnychenko2000field}.

The “beating” or “wave package” visible in the oscillations in Fig.~\ref{fig:4}c have been observed in various experiments \cite{coombs1988fine,kolesnychenko2000field}. Fig.~\ref{fig:6} seems to indicate a number of these wave packages however it is not only the amplitude being modulated also the period of the oscillations shows modulation. We expect on the basis of Fig.~\ref{fig:21} the periods closest to zero to be the largest, with an increasing bias away from zero the periods get smaller. In contrast we observe in Fig.~\ref{fig:6}a a decreasing period getting closer to zero and for Fig.~\ref{fig:6}b an increasing period for voltages increasing towards \SI{8}{\volt}. Both IV curves Fig.~\ref{fig:6}a and Fig.~\ref{fig:6}b show regions where there seems to be destructive interference for a substantial voltage range. The full explanation of this behavior is unclear at present.

\subsection{What is the effect of the partially wet phase?}
This is explained in the framework of the Fermi Golden Rule for tunnel junctions and the more generic form which takes into account energy exchange between tunneling electrons and the environment. The Golden Rule describes a transition rate between an initial state $i$ and a final state $f$ which correlates with the strengths of the coupling between the states $i$ and $f$. The Golden Rule is used successful in a large variety of physical transitions:

\begin{equation}\label{eq:3}
    \vec{\Gamma}_{i\to f}=\frac{2\pi}{\hbar} \left|\mel{i}{Ht}{f}\right|^2 \delta\left(E_f-E_i-\Delta E\right)
\end{equation}

The delta function represents the condition for the energy of the initial and final states: $E_f-E_i=\Delta E$. Applied to the tunnel barrier the Golden Rule states that the transition rate $\vec{\Gamma}_{i\to f}$ represents the transition from an electron in state $i$ on one side of the barrier to state $f$ on the other side of the barrier. For elastic tunneling, where the energy of the electron prior to and after tunneling is identical, equation \ref{eq:3} is described in the following well-known form, at a voltage $V$ for the forward rate, $\vec{\Gamma} (V)$, only:

\begin{equation}\label{eq:4}
    \vec{\Gamma} (V)=\frac{1}{e^2R_T}\int^{\infty}_{-\infty}\dd E f(E)\left[ 1-f(E+eV) \right]
\end{equation}

Usually it is assumed that the electrodes are large on an atomic dimension, to ensure the electrodes contain a large number of free electrons and possess quasi continuous energy spectra. The tunnel resistance $R_T$ is defined by: $R_T=\hbar/(2\pi e^2 \left|\mel{i}{Ht}{f}\right|^2 D^2)$, $D$ represents the density of states in the left and right electrode and is expected to be approximately the same and constant, $f(E)=(1+\exp(E\beta))^{-1}$ is the Fermi Dirac distribution at inverse temperature $\beta=1/k_BT$. In a full quantum mechanical description the tunneling electrons are allowed to interact and exchange energy with the environment. For example the system defining modes related to the charging energy may be excited by exchange of energy quanta $e^2/2C$, just as other environmental modes may be excited by energy quanta absorption from the tunneling electron. Likewise the energy of the tunneling electron can be increased by transfer of energy quanta from the environment to it. Under the 2 conditions that $RT\gg h/2e^2$ (weakly coupled electrodes) and charge equilibrium is established: i.e. the time between tunnel events is larger than the charge relaxation time, equation \ref{eq:4} takes on the following form, for the forward rate, $\vec{\Gamma} (V)$, \cite{devoret1992source4a,ingold1992source4b,devoret1990effect}

\begin{equation}\label{eq:5}
\begin{split}
    \vec{\Gamma} (V)=&\frac{1}{e^2R_T}\int^{\infty}_{-\infty}\dd E \int^{\infty}_{-\infty}\dd E'\big\{f(E)  \\
    &\quad \left[ 1-f(E'+eV) \right]P(E-E')\big\}
\end{split}
\end{equation}

The corresponding physics can be described by a tunneling electron in an initial state with energy E having a multitude of probabilities of tunneling to a multitude of final states, thereby transferring an energy $(E-E')$. If $(E-E')$ is positive the electron transfers energy to the environment, if it is negative the electron absorbs energy from the environment. Quantum mechanics teaches us that we can only know the probability $P(E-E')$ of the tunneling electron ending up in a final state $E'$, thus $\int^{\infty}_{-\infty}P(E)\dd E=1$. For every initial state tunneling to a final state integration over $P(E-E')$ needs to be performed. $P(E)$ is provided for a number of different environments in reference \cite{ingold1992source4b}.

At this stage a “gedankenexperiment” related to the influence of the partially wet phase is postulated. It is assumed that charge is able to accumulate at the THF partially-wet-phase/air interface. The charge is able to freely move at this interface, without being able to flow to the underlying metal conductor. Under these conditions the function $P(E)$ in equation \ref{eq:5} is dramatically impacted since the conducting layer at the partially-wet-phase/air interface effectively prohibits any interaction of tunneling electrons with vacuum environmental modes. The device, enclosed by the partially wet phase layer, provides for the only surviving modes that tunneling electrons can couple to. In other words, the device is caged. A miniature Faraday cage prevents tunneling electrons inside the cage from interacting with low impedance vacuum environmental modes outside of the cage.

\subsection{Continuous Q charging of a THF barrier THF partially-wet-phase junction}
\begin{figure}[H]
    \centering
    \includegraphics[width=.45\textwidth]{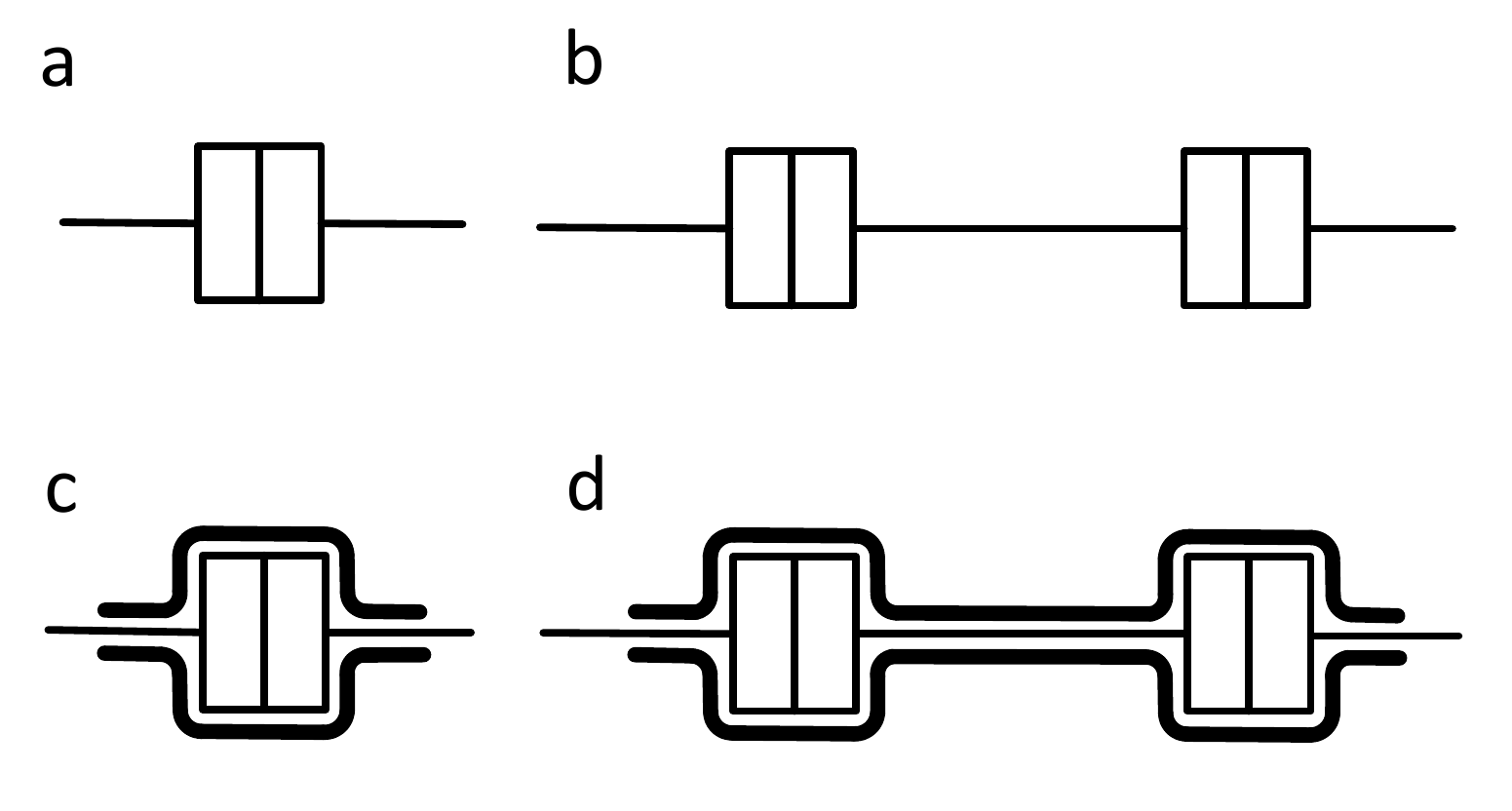}
    \caption{Schematic representation of a: a single junction, b: a double junction, c: a single junction in the partially wet phase, d: a double junction in the partially wet phase.}
    \label{fig:23}
\end{figure}

A THF barrier/partially-wet-phase junction is schematically shown in Fig.~\ref{fig:23}c. The shield lining the junction indicates the junction is in the partially wet phase and thus shielded from vacuum environmental modes.  The well-known unshielded counterpart of this junction is shown in Fig.~\ref{fig:23}a.

A sudden switch to a large voltage bias often leads to a switch on effect as demonstrated in Fig.~\ref{fig:7}a. The current adapts over seconds or 10’s of seconds to the sudden high voltage bias. Other presented graphs may not show this as these graphs may be recorded during continuous scanning after the first “switch on” scan. The switch on effect is attributed to a redistribution of charge at the THF partially-wet-phase/air interface. The partially wet phase layer is finite, it will stop at the cell wall and charge on it will try to counter the electrode charge, acting as a capacitor $C_d$. The reported systems may show a displacement current $I_d=C_d \dv*{V}{t}$, or a continuous increasing current as in Fig.~\ref{fig:9} or decreasing current as in Fig.~\ref{fig:7}b and c as a consequence of the stage the drying-process is in. The resulting drift in the measurements is a given, it does however not prevent the study of the processes taking place inside the junction, which in general correlate to fast timescales in relation to the scale at which the drift takes place.

The charging energy $E_{CH}$ to charge the junction capacitance $C$ with one additional electron on top of the $n-1$ electrons already on the capacitor relates to:

\begin{equation}\label{eq:6}
    E_{CH}(n,Q)=\frac{(Q-ne)^2}{2C},    \quad   n-\tfrac{1}{2}<Q/e<n+\tfrac{1}{2}
\end{equation}

$Q$ is representing the total charge on the capacitor. This equation has been used for the single electron-box decoupled capacitor \cite{introtomes1997introduction} it will in general define a decoupled capacitor. 

\begin{figure}[H]
    \centering
    \includegraphics[width=.49\textwidth]{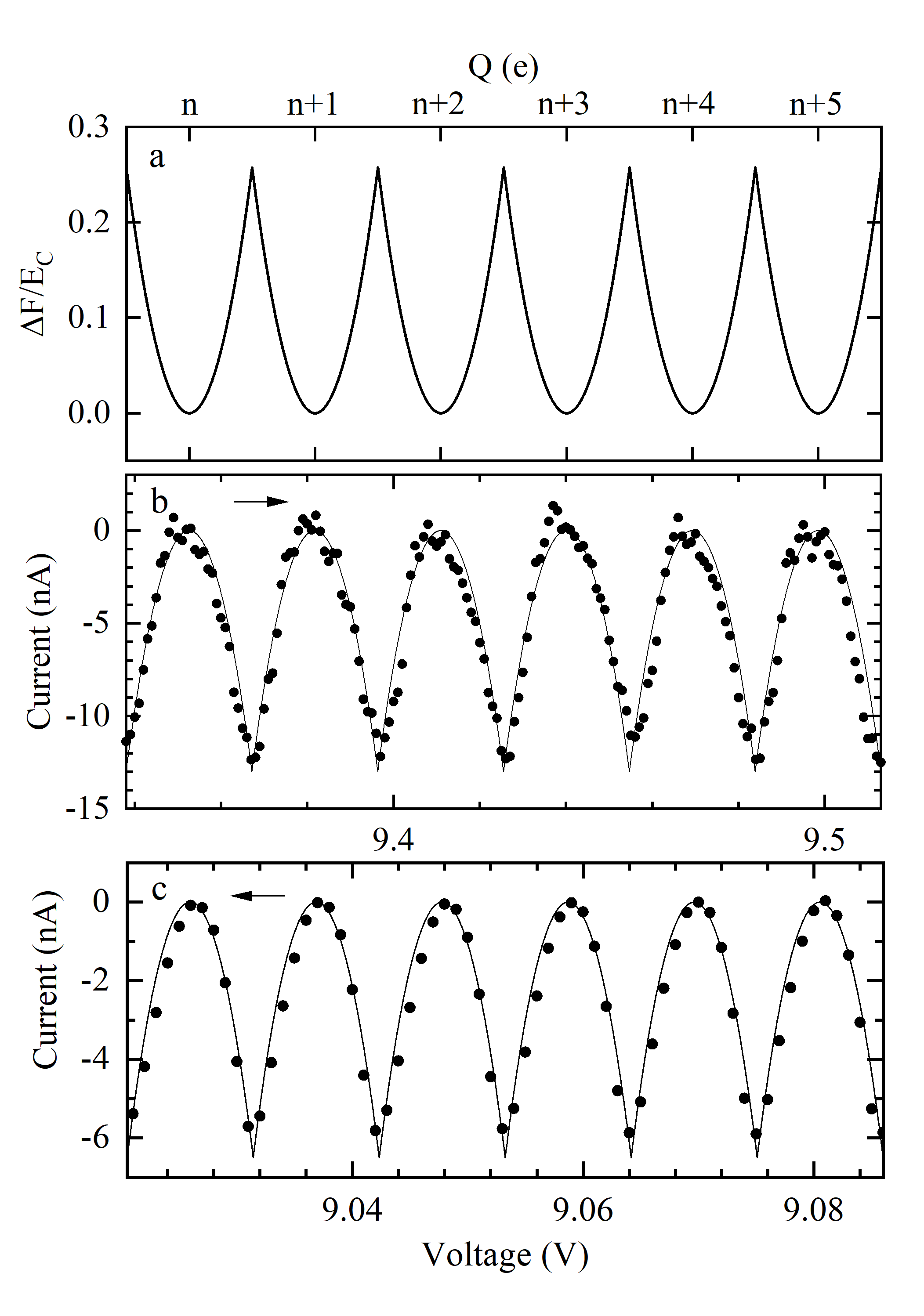}
    \caption{Panel a shows the free energy or the charging energy from equation \ref{eq:6} as a function of $Q$. Panel b and c provide data segments from Fig.~\ref{fig:10} and \ref{fig:7}b respectively. Arrows indicate the scan direction. The continuous line reflects equation \ref{eq:9} with $C=\SI{5.5e-18}{\farad}$, $\alpha=\num{2.3e-3}$ ($I_{ex}=\SI{13}{\nano\ampere}$) for panel b and $C=\SI{1.5e-17}{\farad}$, $\alpha=\num{3.2e-3}$ ($I_{ex}=\SI{6.5}{\nano\ampere}$) for panel c.}
    \label{fig:24}
\end{figure}

The energy $E_{CH}$ is drawn in Fig.~\ref{fig:24}a as a function of $Q$ for various occupation electron numbers $n$. With increasing $Q$, the electron number related to the lowest energy state will increase. This is a continuous process, at the degenerate points there is a change-over to adding an additional electron, only increasing $Q$ at a fractional electron charge. The curve in Fig.~\ref{fig:24}a equals the free energy $\Delta F$ of the system, the energy able to do work. Adding electrons one by one in a continuous manner reveals the nature of the $e/C$ periodicity. 

Only taking into account the contribution from the density of states peak at higher biases ($eV> \Phi^{\text{Au}}$), the current through the molecular junction in a one dimensional conduction channel as indicated in Fig.~\ref{fig:19} follows from equation \ref{eq:3} which modifies to:

\begin{equation}\label{eq:7}
    \vec{\Gamma}_{i\to f}=\frac{2\pi}{\hbar}\left|\mel{i}{Ht}{f}\right|^2\delta\left(E_f-E_i-\Delta F\right)
\end{equation}

Where $\Delta F$ is the free energy catering for the change from state $i$ to state $f$. The matrix element $\mel{i}{Ht}{f}$ ties to both states $i$ and $f$, it represents the strength of the coupling between the two states hence $\left|\mel{i}{Ht}{f}\right|^2$ is proportional to $\Delta F$:

\begin{equation}\label{eq:8}
    \left|\mel{i}{Ht}{f}\right|^2=\alpha\Delta F,  \quad    0<\alpha<1
\end{equation}

It is possible that a number of states constitute an ensemble responsible for the observed quantum effect. For simplicity, transitions between only two states are considered below. Disregarding a linear term in $V$, the current contribution $I_{\text{1D}}=-e\vec{\Gamma}_{i\to f}$ yields:

\begin{equation}\label{eq:9}
\begin{split}
        &I_{\text{1D}}=-\frac{2\pi\alpha e}{\hbar}\frac{(Q-ne)^2}{2C}\\ &n-\tfrac{1}{2}<Q/e<n+\tfrac{1}{2}
\end{split}
\end{equation}

The delta function in equation \ref{eq:7} equals unity for $E_f-E_i=\Delta F$. Equation \ref{eq:9} explains the origin and periodicity of the oscillations shown in Fig.~\ref{fig:7}-\ref{fig:10} it also explains the “V shaped” minima in these oscillations. The sharp minima coincide with the degenerate points at $Q=(n\pm 1/2)e$ of the free energy in Fig.~\ref{fig:24}a. A section of Fig.~\ref{fig:7}b and Fig.~\ref{fig:10} is shown in panel 24c and 24b respectively. The current of these sections have been aligned to the voltage-axis with the top of the oscillations by subtracting a linear function. The capacitance is deduced from the periodicity $\Delta =e/C$, from the current extrema $I_{ex}$, at $(n\pm 1/2)e$, $\alpha$ can be derived. For panel b $I_{ex}=\SI{13}{\nano\ampere}$, $C=\SI{5.5e-18}{\farad}$, $\alpha=\num{2.3e-3}$. For panel c $I_{ex}=\SI{6.5}{\nano\ampere}$ and $C=\SI{1.5e-17}{\farad}$, $\alpha=\num{3.2e-3}$.

The following two processes in this junction are dependent on voltage. The value of Q on the junction electrodes, and thus the free energy $\Delta F(Q)$, can be continuously varied via $Q=CV$. Where the electrodes are in closest proximity tunneling will take place, the peak density of states pinned to the Fermi level will be located at an energy $E=eV$. In scanning the voltage the density of states peak probes the free energy, $\Delta F(Q)$, of the system via coupling to subsequent $Q$ states in a continuous manner. 

The following view emerges for the origin of this quantum effect: the energy $E$ of the tunneling electron is provided to de-populate a filled $Q$ state, to occupy that state while the remaining energy $E'$ is provided to the depopulated electron. At a specific $Q$ state, the energy values for $E$, $E'$, and $\Delta F(Q)$ are each separately identical for every subsequent tunnel event, enabling a coherent time correlated flow of tunneling electrons all sequentially involved in the same processes with the same energy transfers. Due to the existing THF partially wet phase layer, there is no possibility for low impedance vacuum environmental modes to couple into this process, which would break up the coherence and destroy this quantum effect.

In Fig.~\ref{fig:9} a minute feature with a width of about 1-2 \si{\milli\volt} is observed to reproduce in the oscillations. The features appear approximately half way in $Q$-space between the top of the oscillation and the degenerate points. In analogy to the physics of SET multi junction devices this feature is attributed to an excited state in close vicinity of the junction, contributing to the conduction by opening an additional conduction channel. In this case the exited state is coupled via a $Q$ state, being $e/C$ periodic. The coupling takes place via the free energy $\Delta F$ in equation \ref{eq:6}. The excitation energy can be obtained directly by taking $\Delta F$ at the $Q$ state where the excitation is positioned, \SI{0.5}{\milli\electronvolt} in this case. The capacitance related to the oscillations is \SI{1.3e-17}{\farad}, see table \ref{tab1}.

Fig.~\ref{fig:8}c presents a somewhat tilted oscillation as compared to the upright oscillations in Fig.~\ref{fig:8}b. For the increasing scan oscillations tilt to the left, for the decreasing scan a tilt to the right is observed. This change has occurred in the 2 minutes timespan between the two scans. The mechanism behind it is unclear at present. Coupling of the point contact inductance to the charging modes may play a role. From panel 7b a capacitance of \SI{1.5e-17}{\farad} was derived. $E_C$ from this contact equals $\hbar\omega$ with $\omega=(LC)^{-1/2}$ for $L=\SI{3e-10}{H}$. The \SI{e-10}{H} order of magnitude for an inductance related to point-contacts is well known from superconducting rings interrupted by a point-contact \cite{nanohenry1996direct}.

\subsection{Continuous Q charging of a BDT barrier THF partially-wet-phase junction}
An ultimate quantum dot is created, where the BDT conducting phenyl ring is weakly connected to the two electrodes. This double junction is schematically presented in Fig.~\ref{fig:23}d. The shield lining the structure indicates that it is in the partially wet phase and thus shielded from vacuum environmental modes.  The well-known unshielded counterpart of this SET device is shown in Fig.~\ref{fig:23}b.

The characteristic features in the IV curves in Fig.~\ref{fig:11}-\ref{fig:14} are the discontinuities or current jumps. The large periods trailing the discontinuities in Fig.~\ref{fig:11} are rare, observed in only \SI{5}{\percent} of the bending beam assemblies. The smaller periods trailing discontinuities (Fig.~\ref{fig:12}-\ref{fig:14}) are observed in more than half of the bending beam assemblies. Even though the occurrence differs the similarity in shape and slope is clear. Sometimes the periods are too small to clearly attribute a slope to, they appear as oscillations. For those discontinuities where a slope can be attributed it attains without exception a negative differential resistance for increasing voltages and a positive differential resistance for decreasing voltages.

The difference with the THF barrier junction is the conducting island which can only contain an integer number of electrons.  The interesting physics of such a dot is that with an increasing Fermi level of the source electrode, the highest filled level of the dot will in general not be aligned to the Fermi level. The electrodes can obtain any $Q$ value, the dot can only host $n$ electrons, $n$ being an integer. The capacitance as seen from the island consists of a capacitance to the source as well as a capacitance to the drain. Below the total capacitance $C$ as seen from the island is used.

Also for these BDT barrier junctions $e/C$ periodicity is reflected. For devices where the single electron levels are small compared to $e^2/C$ the spin degeneracy of the levels is lifted and the charging energy regulates the spacing \cite{beenakker1991granular}. With an increasing $E_F$ at some point an electron occupies an additional single electron level on the dot with an energy $e^2/C$ above the previous level. During the increase of $E_F$ and prior to the “electron to island tunneling” the free energy $\Delta F$ increases as well. The Fermi level increased above the last filled level in the dot. An imbalance is created in the energy spectrum of the dot, less than $e^2/C$. As a consequence $\Delta F$ increases with this amount. This lasts until the free energy can be reduced by the amount $e^2/C$, at that stage an electron tunnels to the dot. By reducing $E_F$ the situation is not invariant as also in this case the free energy is increased. In this case $E_F$ is below the highest filled dot level. When $E_F$ gets $e^2/C$ below it an electron leaves the dot. At that point the free energy can be reduced by $e^2/C$ which it gained in this cycle. 

\begin{figure}[H]
    \centering
    \includegraphics[width=.49\textwidth]{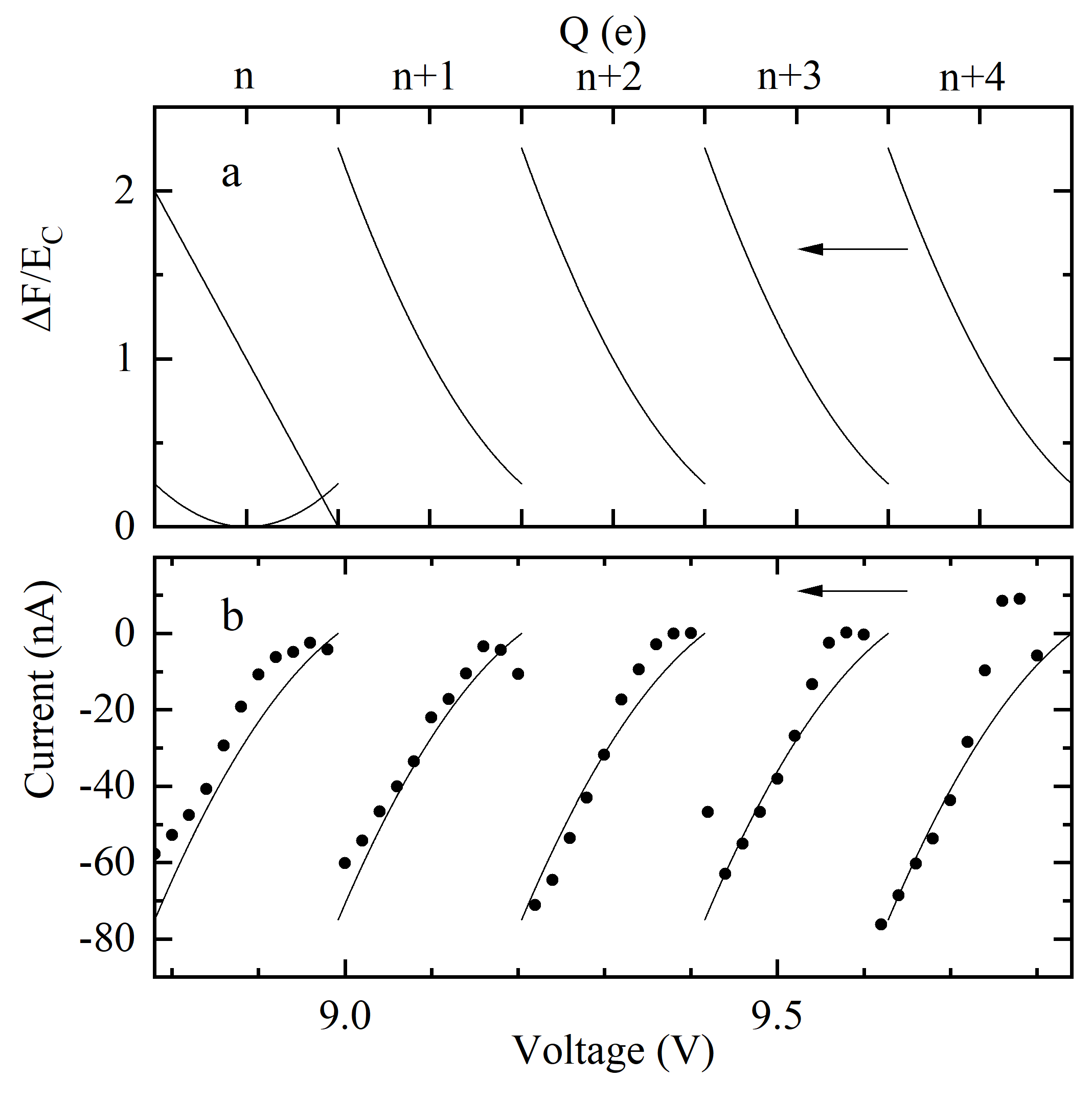}
    \caption{In panel a the two contributions to the free energy, charging energy and charge imbalance, are indicated to the far left as a function of $Q$. The remaining curves express the combined contribution. Data from a segment of Fig.~\ref{fig:11}a are shown in panel b as well as the inverted and scaled free energy curves after aligning to zero. Arrows indicate the scan direction.}
    \label{fig:25}
\end{figure}

\begin{figure}[H]
    \centering
    \includegraphics[width=.49\textwidth]{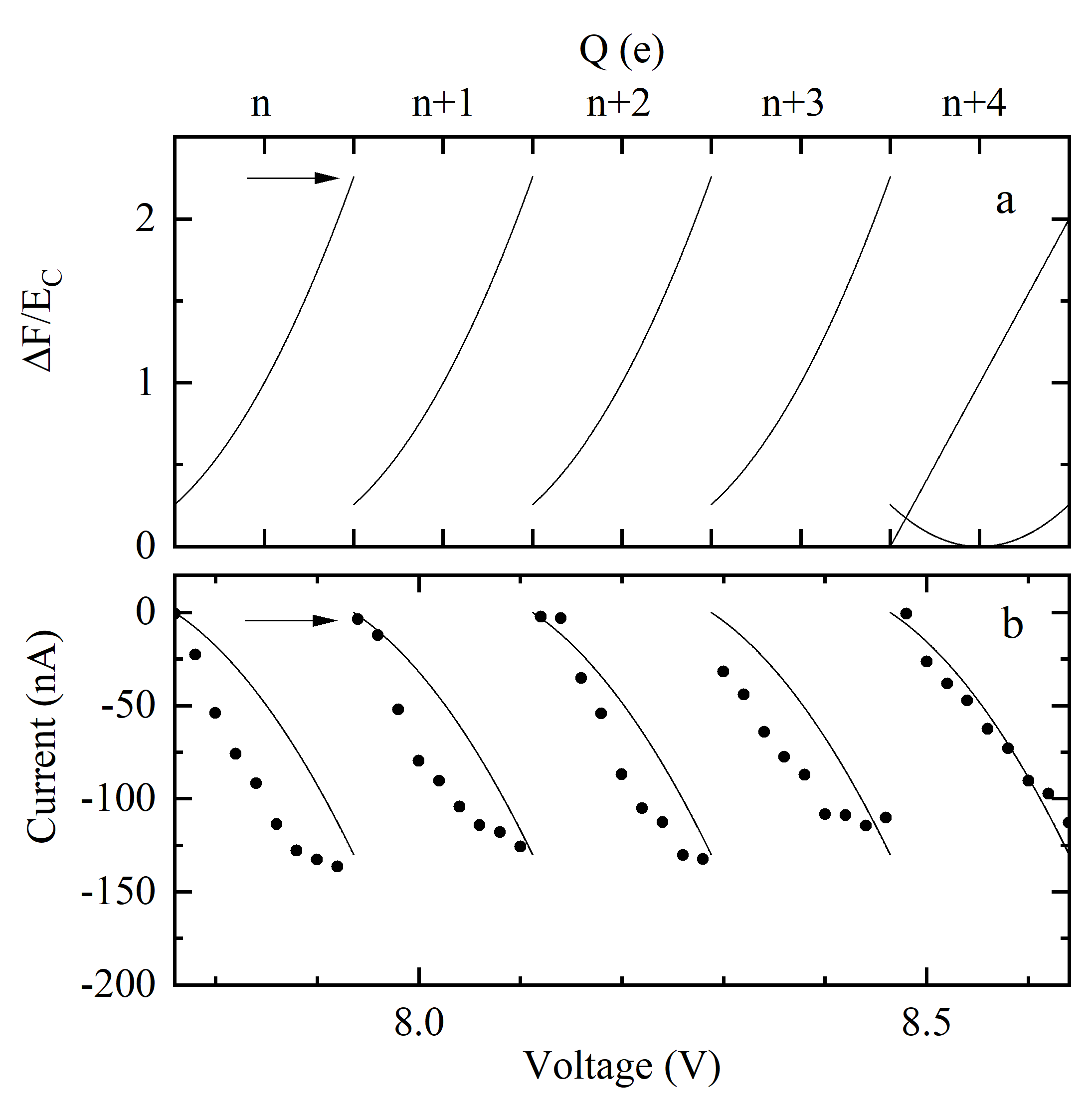}
    \caption{Panel a shows the two separate contributions to the free energy at the far right as a function of $Q$. Data from a segment of Fig.~\ref{fig:11}b are shown in panel b as well as the inverted and scaled free energy curves after aligning to zero. Arrows indicate the scan direction.}
    \label{fig:26}
\end{figure}

This process is detailed in Fig.~\ref{fig:25} and Fig.~\ref{fig:26} where in panel a the free energy is provided for the charging of the capacitor as well as for the charge imbalance on the dot for both increasing and decreasing $E_F$. Both contributions are separately indicated to the far left in Fig.~\ref{fig:25}a and the far right in Fig.~\ref{fig:26}a. The remaining curves present the combined contribution to the free energy. A section of Fig.~\ref{fig:11} a and b is shown in Fig.~\ref{fig:25}b and Fig.~\ref{fig:26}b respectively, the data have been aligned to the voltage-axis with the top of the periodic structure by subtracting a linear function. The positive and negative differential resistance parts in the IV discontinuities naturally find their origin from detailing the total free energy $\Delta F$, which increases saw tooth wise both for increasing and decreasing voltage scans. The minimum of $\Delta F$ was set at 0. The magnitude of $\Delta F$ is $2 E_C$. The function was scaled to match the current extrema in Fig.~\ref{fig:25}c and Fig.~\ref{fig:26}c.

The capacitance is derived from the $e/C$ periodicity, the current extrema are obtained from the data providing: $I_{ex}=\SI{75}{\nano\ampere}$, $C=\SI{7.5e-19}{\farad}$ and $I_{ex}=\SI{130}{\nano\ampere}$, $C=\SI{9.1e-19}{\farad}$ for Fig.~\ref{fig:25} and Fig.~\ref{fig:26} respectively. The equations in section 4.4 are derived for a single junction, caution is required interpreting a derived $\alpha$ for a double junction system.

Based on the free energy it is established that every $e^2/C$ cycle an electron tunnels to or from the island depending on an increasing or decreasing $E_F$, since the free energy can be lowered by this energy amount. The resulting discontinuities are observed in the IV curves as current increases. In Fig.~\ref{fig:14} for example over 100 electrons are tunneling in individual $e/C$ voltage cycles to the island for an increasing voltage in the range from \SIrange{9}{10}{\volt}. Why is the level spacing of the single electron levels so small in a molecule where we expect \si{\electronvolt} levels rather than \si{\milli\electronvolt} levels? At this stage this is not clear. It is possible that the interaction of the BDT molecule with the lining THF molecules leads to many more levels as compared to the original BDT molecule.

The data in Fig.~\ref{fig:15} show many similarities with the smooth THF barrier THF partially-wet-phase junction oscillations with sharp “V-shaped” minima. No discontinuous current jumps are visible in panel b. Even though the BDT molecule solution has been used, it shows the unsuccessful attempt to create a quantum dot. It also shows that in being unsuccessful, most likely a THF barrier THF partially-wet-phase junction has been created instead. It is possible that not the entire electrode surface has been covered with BDT molecules or that some molecules went down with both thiol groups on one electrode, the resulting IV curve is very similar to the THF barrier THF partially wet phase junctions. Panel b and c show a shoulder in the oscillations appearing and increasingly causing the regular oscillation to break up. Similarly to Fig.~\ref{fig:9} this is attributed to excited states coupled to a $Q$ state, entering the $e/C$ window.

Fig.~\ref{fig:16} provides insight in the criticality of the partially wet phase. The IV curve shows a somewhat irregular structure, for the increasing voltage scan the structure shows at least for a number of discontinuities a negative differential resistance. Electrons are tunneling onto the dot. In the decreasing voltage scan electrons are tunneling from the dot. This effect disappears abruptly once the partially wet phase has disappeared in the lower curve at \SI{9.25}{\volt}. What happens to the hundreds of electrons that may be stored on the dot at this point? The nature of the quantum effect disappears once vacuum environmental modes start to mingle in the conduction. The conduction becomes incoherent single electron effects are no longer visible. The hundreds of electrons may remain on the dot, however the intricate and subtle difference of adding or subtracting them one by one, gets lost.

\subsection{Capacitance reviewed}
This section is deviating from the electrostatic capacitance assumption. Defining capacitance at the microscopic level is complex. In one view \cite{buttiker1986traversal} the characteristic time $\tau_t$ the electron takes to traverse the barrier while tunneling is determining the distance $\tau_tc$ it surveys, c is the charge propagation speed. Nazarov posed a different view \cite{nazarov1989anomalies} where the time $\tau_N$, defined by $\tau_N=\hbar/eV$, is the relevant parameter for the junction-lead system. This leads to a distance $\tau_Nc$, being surveyed by the electron. The high bias voltages up to \SI{10}{\volt} in the shown junction results will lead to extremely small $\tau_Nc$ distances as compared to normal measurements in the sub \SI{0.1}{\volt} range.

\begin{figure}[H]
    \centering
    \includegraphics[width=.49\textwidth]{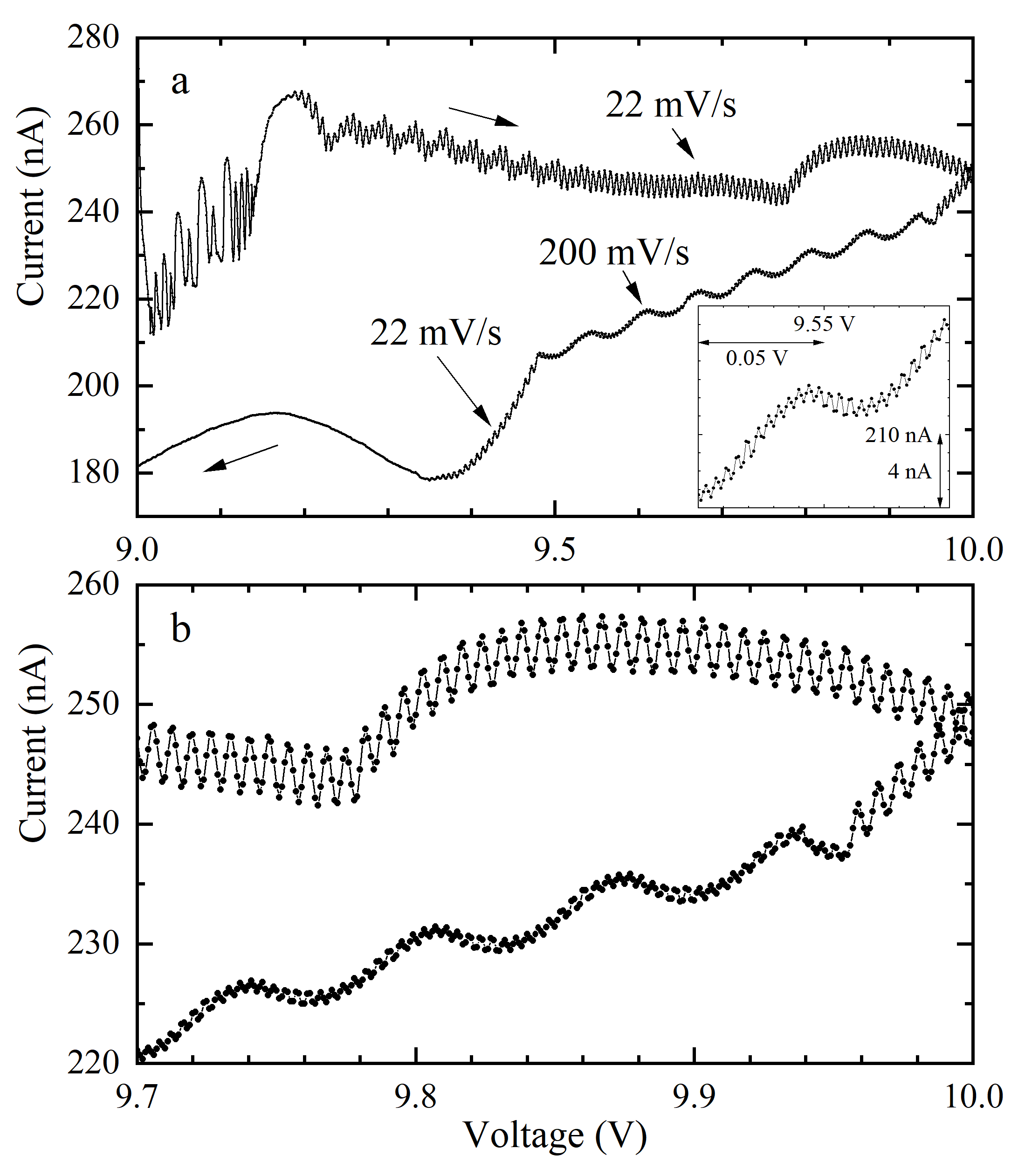}
    \caption{A THF barrier junction in the partially wet phase where during one segment of the IV curve the voltage scan rate increased from \SI{22}{\milli\volt\per\second} to \SI{200}{\milli\volt\per\second}. The periods of \SI{7.1}{\milli\volt} are consequently enlarged to \SI{66.6}{\milli\volt} while the amplitude remains similar. In addition a small oscillation is present on the large period oscillation. This is also shown in the inset of panel a. Panel b is an enlargement of panel a, Au\textsuperscript{S}, $\phi_{\text{in}}^{2}$.}
    \label{fig:27}
\end{figure}

\begin{figure}[H]
    \centering
    \includegraphics[width=.49\textwidth]{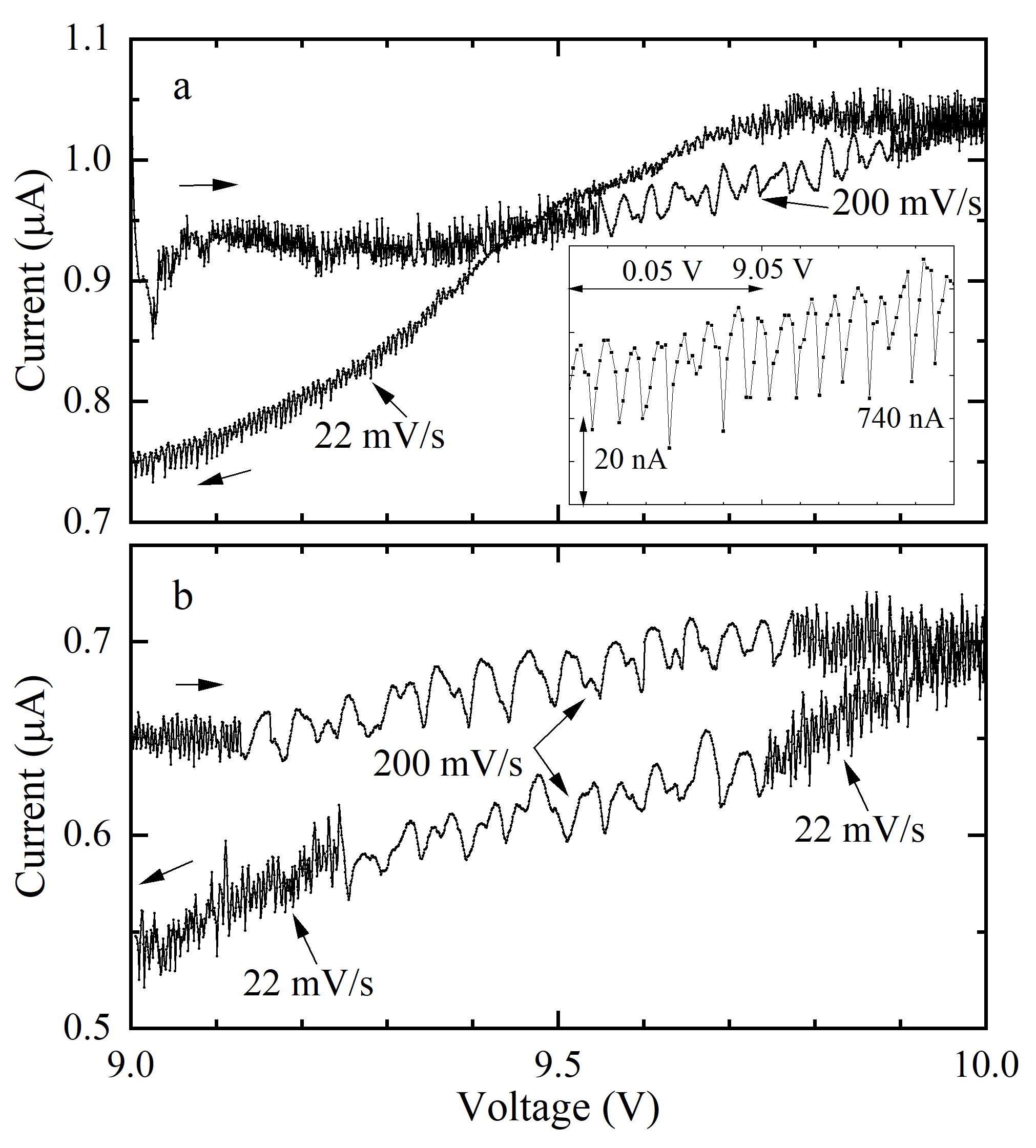}
    \caption{A THF barrier junction in the partially wet phase. Panel a and b show two IV traces recorded after each other where for segments of the traces the scan-speed was enlarged. The inset of panel a shows an enlargement of a 0.1 voltage section, Au\textsuperscript{S}, $\phi_{\text{in}}^{2}$}
    \label{fig:28}
\end{figure}

Fig.~\ref{fig:27} and Fig.~\ref{fig:28} show data of two bending beam assemblies THF barrier junctions in the partially wet phase. In these IV curves the scan speed was modified from the standard fairly low \SIrange{22}{200}{\milli\volt\per\second} for a few indicated voltage sections. Fig.~\ref{fig:27}b is an enlargement of 27a, the IV curve in Fig.~\ref{fig:28}b was recorded just after the curve in Fig.~\ref{fig:28}a. In Fig.~\ref{fig:27} oscillations in the \SI{22}{\milli\volt\per\second} sections are visible. Increasing the scan-speed 9 fold is increasing the period of the oscillations by a factor of 9 as well. For Fig.~\ref{fig:28} a similar behavior is shown. In this figure the black parts appear noisy, however by zooming in it is clear that this is reflecting small oscillations, see for example the inset. The 9 fold increase in scan-speed also leads to an approximate 9 fold increase in period in both panels a and b, where the period of the smaller oscillations has been measured next to the increased scan-rate section. It is notable that the amplitude of the large period oscillations remains very similar to the amplitude of the original small period oscillations. The above described behavior on scan-speed change and amplitude has also been measured in BDT barrier junctions in the partially wet phase.

\begin{figure}[H]
    \centering
    \includegraphics[width=.49\textwidth]{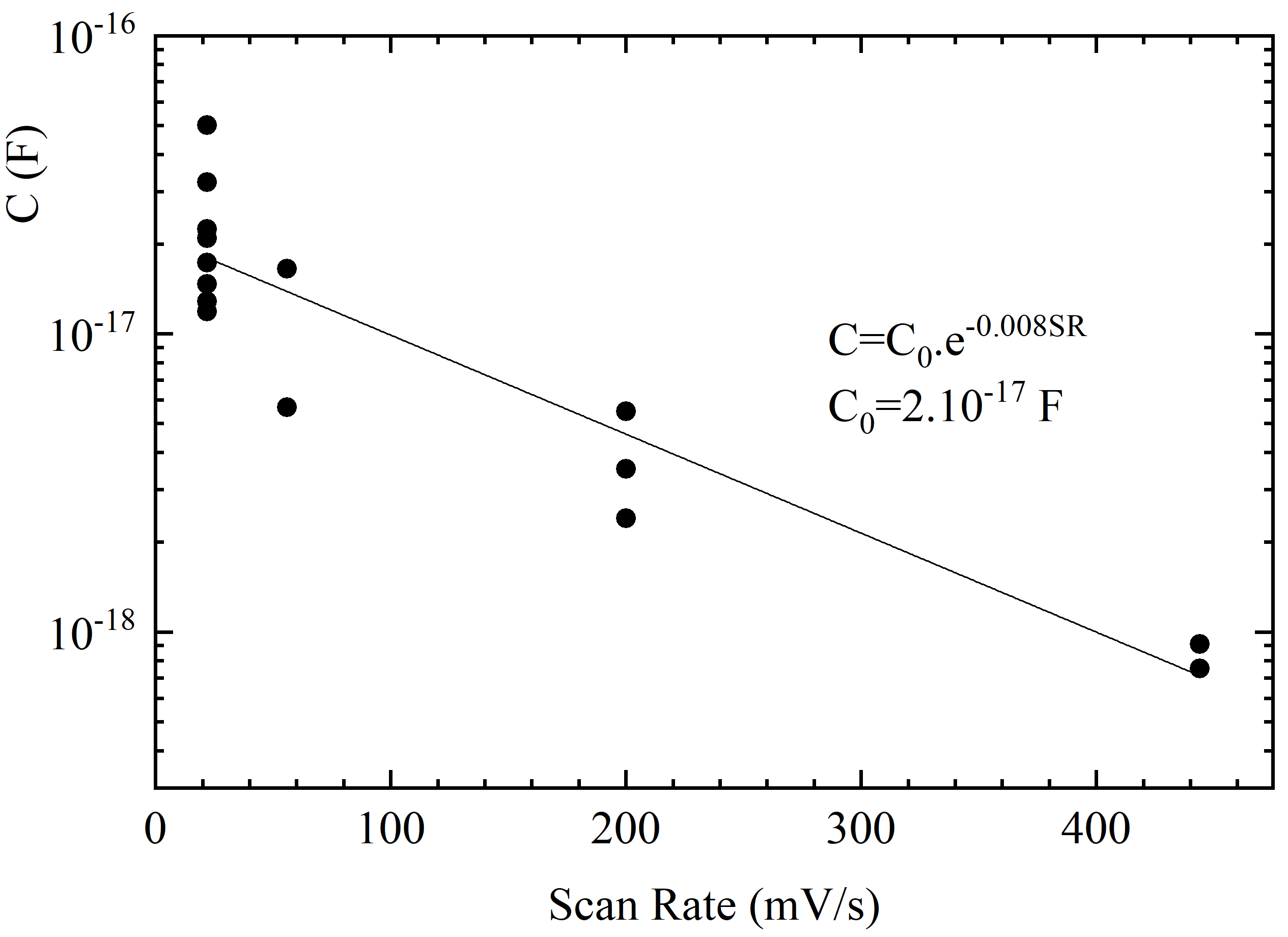}
    \caption{Capacitance values from table 1 as a function of the used voltage scan-speed in the measurement. The line represents an exponential fit to the data points; $C=C_0\exp(-8\text{SR})$ with SR the scan-rate in \si{\volt\per\second} and $C_0=\SI{2e-17}{\farad}$.}
    \label{fig:29}
\end{figure}

Table 1 summarizes the results from Fig.~\ref{fig:27} and Fig.~\ref{fig:28} as well as from all other figures where periodic oscillations have been measured. The capacitance has been derived from the periodicity $e/C$. Fig.~\ref{fig:29} shows the derived capacitance as a function of the used scan rate in the experiment. This graph shows that by increasing the scan-rate by a factor of 20 the capacitance is reduced by a factor of 20, consistent with the findings in  Fig.~\ref{fig:27} and Fig.~\ref{fig:28} where the period increased by a similar fraction as the scan-rate increase.

Practically the capacitance can be described by $C_0\exp(-8\text{SR})$ with SR the scan-rate in \si{\volt\per\second} and $C_0=\SI{2e-17}{\farad}$. The data in the preceding sections are still valid, however now it is known that it would have been different with a different scan-rate.

Essentially the scan-rate of \SI{22}{\milli\volt\per\second} is very low and close to the $C_0$ value in a static situation. The dependence of period and capacitance on scan-rate may also explain the rare occurrence of curves as indicated in Fig.~\ref{fig:11} which were measured at the highest scan-rate.

\begin{table}[H]
\centering
\caption{Summarizing capacitance values and scan-rates of periodicity reflected in the various Figures. The last point is added and not reported in any of the Figures.}
\label{tab1}
\begin{tabular}{@{}lllll@{}}
\toprule
Figure & Period & Capacitance & Scan Rate & $\dv*{Q}{t}$ \\
&\multicolumn{1}{c}{(\si{\milli\volt})} & \multicolumn{1}{c}{(\si{\farad})} & \multicolumn{1}{c}{(\si{\milli\volt\per\second})} & \multicolumn{1}{c}{(\si{e\per\second})}\\
\midrule
\ref{fig:10}           & 29.2  & \num{5.5e-18} & 200 & 6.9 \\
\ref{fig:7}b           & 10.9  & \num{1.5e-17} & 22  & 2.0 \\
\ref{fig:11}a          & 212.0 & \num{7.5e-19} & 444 & 2.1 \\
\ref{fig:11}b          & 176.0 & \num{9.1e-19} & 444 & 2.5 \\
\ref{fig:9}            & 12.5  & \num{1.3e-17} & 22  & 1.8 \\
\ref{fig:12}           & 13.5  & \num{1.2e-17} & 22  & 1.6 \\
\ref{fig:13}ab         & 9.2   & \num{1.7e-17} & 22  & 2.4 \\
\ref{fig:13}cd         & 7.6   & \num{2.1e-17} & 22  & 2.9 \\
\ref{fig:15}bc         & 28.2  & \num{5.7e-18} & 56  & 2.0 \\
\ref{fig:27}           & 66.6  & \num{2.4e-18} & 200 & 3.0 \\
\ref{fig:27}           & 7.1   & \num{2.2e-17} & 22  & 3.1 \\
\ref{fig:28}a          & 3.2   & \num{5.0e-17} & 22  & 6.9 \\
\ref{fig:28}a          & 29.1  & \num{5.5e-18} & 200 & 6.9 \\
\ref{fig:28}b          & 45.4  & \num{3.5e-18} & 200 & 4.4 \\
\ref{fig:28}b          & 5.0   & \num{3.2e-17} & 22  & 4.4 \\
not rep. & 9.7   & \num{1.6e-17} & 56  & 5.8 \\ \bottomrule
\end{tabular}
\end{table}

It appears that increasing the scan-rate reduces the $\tau_Nc$ distance, the horizon of the electron. This is reducing the capacitance. The scan-rate increase, coinciding with a proportional capacitance reduction, leaves $\dv*{Q}{t}$ constant, within the accuracy of the experiment. Fig.~\ref{fig:29} shows $\dv*{Q}{t}$ for the junctions indicated in table 1, the data points from Fig.~\ref{fig:27} and Fig.~\ref{fig:28} are pairwise highlighted, indicating that $\dv*{Q}{t}$ remains constant at a scan-rate increase. The amplitude of the oscillations remains constant at a scan-speed change. The amplitude correlates to the free energy in equation \ref{eq:8} which in turn correlates with the charging energy in equation \ref{eq:6}. The capacitance behavior at scan-speed changes related to the periodicity appears to be different from the capacitance behavior related to the free energy scale.

Summarizing this section, the high voltage is plausible to play a role as to why single junctions are decoupled from the leads and display small capacitance effects. The measured periodicity/capacitance is dependent on the scan-rate. At scan-speed changes $\dv*{Q}{t}$ remains constant.

\begin{figure}[H]
    \centering
    \includegraphics[width=.49\textwidth]{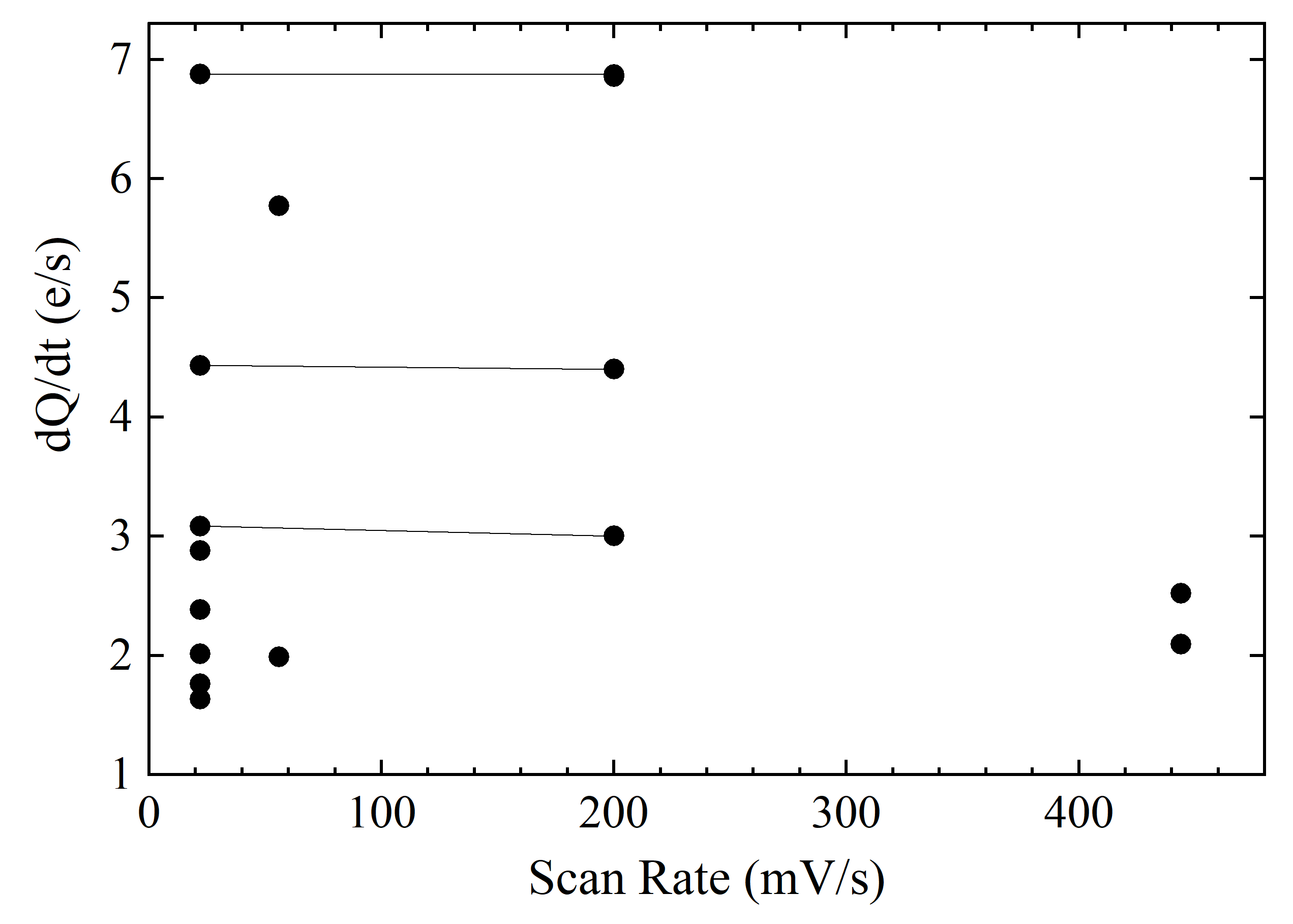}
    \caption{$\dv*{Q}{t}$ as a function of scan-speed for the various junctions indicated in table 1. The data points from the three IV curves in Fig.~\ref{fig:27} and Fig.~\ref{fig:28} are pairwise highlighted as they originate from the same IV curve.}
    \label{fig:30}
\end{figure}

\subsection{General discussion}
One of the striking observations is related to temperature. Why is it possible to observe effects in a single junction at room temperature where single electron tunneling is regularly studied at dilution fridge temperatures? Why is it possible to reproduce a 1-2 \si{\milli\volt} feature, measure sharp \si{\milli\volt} width “V-shaped” curves at room temperature where $k_BT$ equals \SI{25}{\milli\electronvolt}? The answers to these questions will probably have the same basis as the existence of this quantum effect. One aspect will be that the tunneling electrons are simply not capable of interacting with any other modes than a few device defining modes.

\begin{figure}[H]
    \centering
    \includegraphics[width=.49\textwidth]{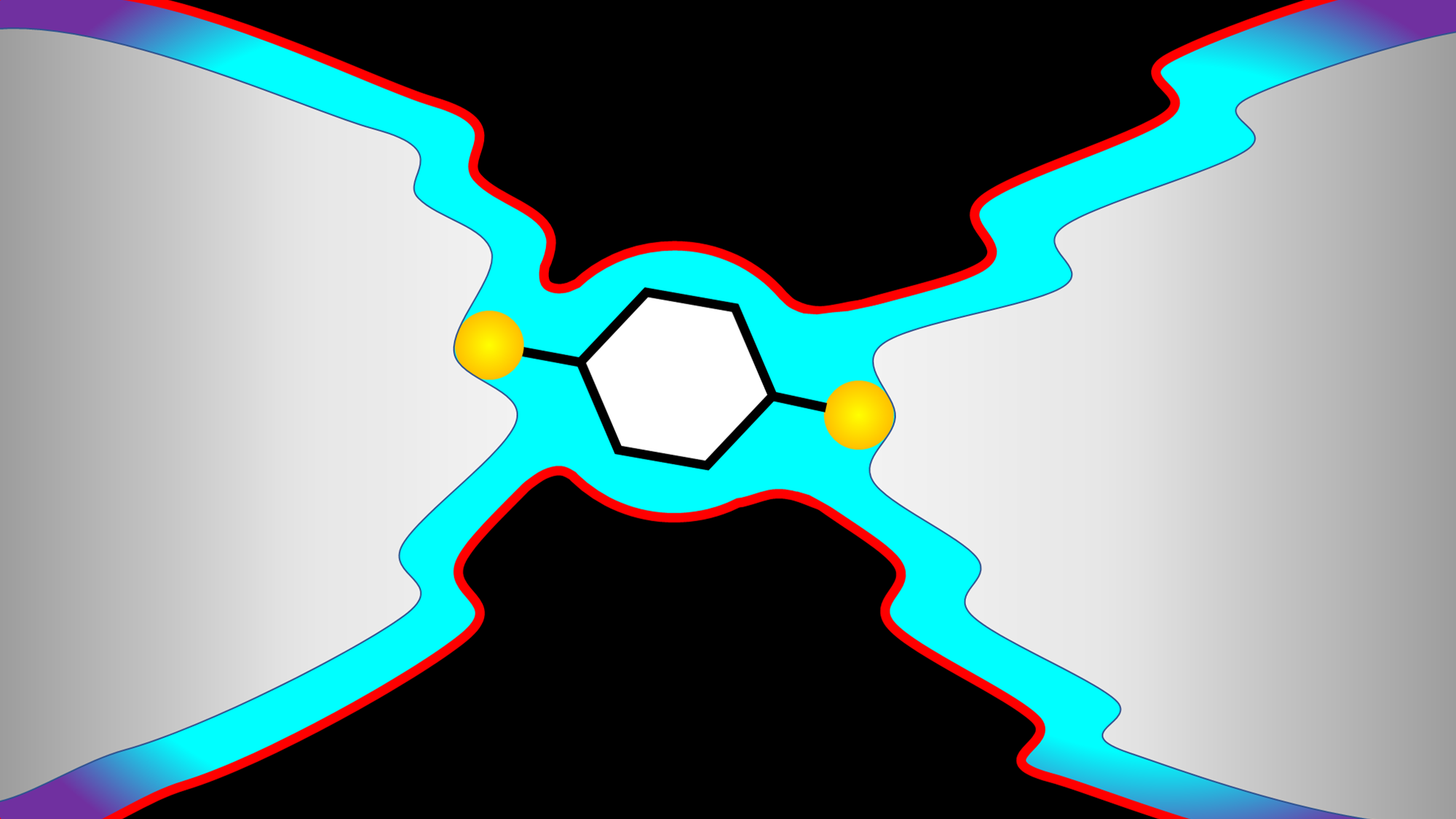}
    \caption{An impression of the BDT barrier junction in the partially wet phase, not to scale. The blue/purple layer represents the THF partially wet phase layer the red liner represents the finite width of the surface potential trapping charge carriers at the THF/air interface.}
    \label{fig:31}
\end{figure}

Confining charge carriers to a plane is not new; the two dimensional electron gas (2DEG) has advanced our understanding of science over the past 4 decades. Bandgap engineering in a GaAs-AlGaAs heterojunction can create a potential trap for electrons in the surface potential perpendicular to the interface leading to a 2DEG at the interface of the electronically different materials GaAs and AlGaAs. Here, nature provides for a perfect conductive layer, lining the molecular device via the THF partially wet phase. In this case the charge gas follows a contour-shaped liner layer. The surface potential-trap is perpendicular to the THF partially wet phase layer/air interface, see Fig.~\ref{fig:31}. The trapped charge carriers are free to move in the plane of the lining layer, in the same way as in a 2DEG. Based on the presented experiments, the fact that the partially wet phase is a requirement to be able to observe these effects, the analogy to the physical principle of a 2DEG at material interfaces, as well as the explanation and derivation of the oscillations and line shapes in the IV curves, the only possible conclusion is that the earlier posed “gedankenexperiment” is in actual fact the reality.

The above does not imply that every partially wet phase layer will have floating charge carriers, in fact for H$_2$O it was shown that the partially wet phase layer is able to hold charge at the electrode side captive due to the dipole moment of the H$_2$O molecule. It is simply not known at this stage how charge at the H$_2$O/air side will behave. 

The high voltage and the THF partially wet phase seem to be a combined requirement for the observation of the reported quantum effects in the THF barrier and BDT barrier junctions. There are multiple experimental findings revealing that the $Q$-states are real: $\dv*{Q}{t}$ has been shown to be constant for a junction. The IV line shape of the THF barrier junction shows behavior as predicted by the free energy. Excited states couple to specific $Q$-states and reproduce over the oscillations with $e/C$ periodicity. The BDT barrier junction shows expected behavior with respect to tunneling to and from an island depending on scan direction. The voltage scan-rate couples to the inner quantum mechanics of the junction, leading to scan-rate dependent IV curves. Based on the above as well as the combined presented results we feel confident stating that the $Q$-side of the commutator has been reached. This implies that the Schön phase, $\varphi_S$, has become undefined and $Q$ needs to be treated as a classical variable.

The effects of the partially wet phase on fundamental tunnel properties are probably not restricted to molecular junctions only. Small capacitance solid state tunnel junctions and compiled multi-junction systems, lined with a single molecular layer of a well-chosen gas, at cryogenic/dilution-fridge temperatures are potentially interesting devices. Almost half a century ago Kulik and Shekhter predicted their voltage oscillations \cite{kulik1975kinetic}. We are now on the verge of discovering the full potential of these quantum effects.

In the end Molecular Electronics will be unified with the ongoing device miniaturization by the partially wet phase.

\section{Conclusions}
A method to produce molecular junctions in the partially wet phase has been presented. In the partially wet phase the conduction of these systems is essentially one dimensional. The connection to the “known” has been established via observed Grundlach oscillations. Reproducible and long-term enduring single molecular junctions at ambient conditions can be realized. At high voltages a quantum effect related to continuous $Q$ charging has been demonstrated. The THF barrier junction reflects $e/C$-periodicity of the free energy. In the BDT barrier junction IV curve discontinuities reflect electron tunneling to or from an island. For a particular junction $\dv*{Q}{t}$ is constant and independent of the voltage scan-rate. $Q$ needs to be treated as a classical variable.

The final conclusion is that the THF partially wet phase is responsible for shielding interactions between tunneling electrons and vacuum environmental modes.

\section*{Acknowledgements}
We would like to thank F.L. Muller for many stimulating, encouraging and constructive discussions.

Not because it was easy, but because he had a conviction it was worthwhile pursuing, von Hippel had a visionary belief in molecular electronics. Up to this date we are discovering new physics in this field, progressing von Hippel’s chess game to an interesting next phase.


\bibliographystyle{unsrt}
\bibliography{references}

\end{multicols}

\end{document}